\documentclass[%
 reprint,
 amsmath,amssymb,
 aps,
prb,
]{revtex4-2}
\usepackage{braket}
\usepackage{comment}

\usepackage{graphicx}
\usepackage{tipa}
\usepackage{amsfonts}
\usepackage{bbding}
\usepackage{bbold}
\usepackage{bbm}
\usepackage[T1]{fontenc}
\usepackage{amsmath}
\usepackage{color}
\usepackage{soul}
\usepackage{ulem}
\usepackage{bbm} 
\usepackage{tensor}
\usepackage{dsfont}
\usepackage{slashed}
\usepackage{hyperref}
\usepackage[all]{hypcap}
\hypersetup{colorlinks=true,
   linkcolor=blue,
    citecolor=blue,
    filecolor=magenta,
    urlcolor=cyan}
  
\begin{document}

\title{Magnetic phase transitions protected by topological quantum geometry transitions: effects of electron-electron interactions in the Creutz ladder system}

\author{Abdiel de Jesús Espinosa-Champo$^{a,c}$}
\author{Gerardo G. Naumis$^{b}$}
\email[e-mail: ]{naumis@fisica.unam.mx}

\affiliation{${}^{a}$Posgrado de Ciencias F\'isicas, Universidad Nacional Autónoma de México, Apartado Postal 20-364 01000, Ciudad de México, México.}
\affiliation{${}^{b}$Depto. de Sistemas Complejos, Instituto de F\'isica, Universidad Nacional Aut\'onoma de M\'exico (UNAM). Apdo. Postal 20-364, 01000, CDMX, Mexico.}
 \affiliation{${}^{c}$ Departamento de F\'isica, Facultad de Ciencias, Universidad Nacional Aut\'onoma de M\'exico, Apdo. Postal 70-542, 04510, CDMX, México.}

\begin{abstract}

The interplay between electronic correlations and band topology is a central theme in modern condensed matter physics. In this work, we investigate the effects of on-site Hubbard interactions on the topological, magnetic, and quantum geometric properties of the Creutz ladder, a paradigmatic model of a one-dimensional topological insulator. Using a self-consistent mean-field approach, we uncover a first-order, interaction-driven phase transition that is simultaneously magnetic and topological. We demonstrate that as the Hubbard interaction $U$ is increased, the system's ground state abruptly switches from an anti-ferromagnetic (AF) configuration to a ferromagnetic (F) one. This magnetic transition coincides with a topological transition, marked by a quantized jump in the Zak phase from $\pm\pi$ to $0$. We systematically compute the phase diagrams in the parameter space of on-site energy staggering ($\epsilon$) and inter-chain hopping asymmetry ($\lambda$), revealing the critical interaction strength $U_c$. Furthermore, we analyze the quantum geometry of the Bloch states by calculating the Fubini-Study metric, demonstrating that its components exhibit divergences that precisely signal the topological phase transition. By analyzing the full energy spectrum, we distinguish the true ground state from metastable excited states that emerge past the critical point. Our results establish the Creutz-Hubbard ladder as a minimal model for studying interaction-induced topological phenomena and suggest a potential route for controlling magnetic, topological, and geometric properties via electronic correlations.

\end{abstract}

\maketitle

\section{Introduction} \label{sec: Introduction}

The intersection of electronic correlations and non-trivial band topology represents one of the most active and fruitful frontiers in modern condensed matter physics \cite{Hasan2010_Review, Tokura2019_Review, Wen2017_Review}. While topological band theory has successfully classified non-interacting systems—leading to the discovery of topological materials \cite{Bernevig2013_Book} like topological insulators \cite{Kane2005_QSH, Bernevig2006_QSHG, Espinosa-Champo_2024_Flat_bands,espinosachampo2025hyperbolicplasmondispersionoptical}, Dirac semimetals \cite{Wang2012_Dirac, Young2012_Dirac,borophene_abdiel}, and other exotic phases of matter \cite{Chiu2016_Review, Armitage2018_Weyl,multifractal_abdiel}—the role of strong electron-electron interactions remains a profound and challenging open question. The Hubbard model, a cornerstone for describing correlated electron systems, provides the canonical framework for phenomena such as Mott insulators and magnetism \cite{Arovas2022_HubbardReview, Imada1998_Mott}. The interplay between these two paradigms, topology and correlation, is predicted to spawn a host of novel quantum phenomena, including topological Mott insulators \cite{Pesin2010_Mott, Rachel2010_Mott}, fractional Chern insulators \cite{Neupert2011_FCI, Regnault2011_FCI}, and interaction-driven topological phase transitions \cite{Varney2010_InteractionTPT, Budich2012_InteractionTPT, espinosachampo2025adiabaticitystudytopologicalphases}.

One-dimensional lattice models serve as ideal platforms for exploring this interplay due to their analytical and numerical tractability. Among these, the Creutz ladder stands out as a paradigmatic model for one-dimensional topological insulators \cite{Creutz1999}. Even in its non-interacting form, the Creutz ladder exhibits remarkable features, including the presence of perfectly flat, dispersionless bands \cite{Leykam2018_FlatBandReview, EspinosaChampo2023_Fubini} and a quantized topological invariant—the winding number \cite{Ryu2010_Classification} or Zak phase \cite{Zak1989}. These properties, which have been realized experimentally in synthetic systems like photonic lattices \cite{Kang_2020,He2021,Meng2016,Rechtsman2013_Photonic, Mukherjee2017_PhotonicCreutz} and cold atoms \cite{Piga2017,Kuno_2020,Atala2013_ColdAtom, Aidelsburger2013_ColdAtom}, make it an exceptional candidate for investigating how robust topological features are in the presence of strong electronic interactions and how these interactions might, in turn, induce new topological states \cite{Zurita2020,Piga2017, Varney2010_InteractionTPT, Kang_2020}.
The role of electronic correlations in the emergence of topological phenomena is a subject of intense research. In this context, recent work by Bouzerar \textit{et al.} \cite{Bouzerar2025} has explored the spin dynamics in the Hubbard-Creutz ladder, revealing that correlations can induce spin-polarized phases. Our study aligns with this approach, but delves deeper into the nature of the interaction-induced phase transitions. Specifically, we show that an increase in the interaction $U$ not only generates a spin-polarized phase, but also induces an abrupt transition from an anti-ferromagnetic (AF) to a ferromagnetic (F) configuration, which is intrinsically tied to a quantized topological transition. Furthermore, we extend the analysis to include the quantum geometry of the Bloch states, demonstrating that this topological and magnetic transition leaves a distinct signature in the Fubini-Study metric, which exhibits a precise divergence at the transition point.

The central question we address is how the on-site Hubbard interaction, $U$, modifies the electronic and magnetic ground state of the Creutz ladder. While the non-interacting model can be tuned between trivial and topological phases via hopping parameters, introducing correlations opens the door to spontaneous symmetry breaking and the emergence of ordered phases. Specifically, the competition between inter-site hopping and on-site repulsion can lead to complex magnetic ordering, such as anti-ferromagnetism (AFM) \cite{Neel1948_Antiferromagnetism} or ferromagnetism (FM) \cite{Stoner1938_Ferromagnetism}. Understanding the connection between these emergent magnetic phases and the underlying band topology is crucial for engineering new functionalities in quantum materials.

In this work, we investigate the phase diagram of the Hubbard model on a Creutz ladder using a self-consistent mean-field approach. We demonstrate the existence of an interaction-driven phase transition that is simultaneously magnetic and topological. Our numerical results reveal a first-order phase transition where, by increasing the Hubbard interaction $U$, the system's ground state switches from an anti-ferromagnetic (AF) state to a ferromagnetic (F) state. Remarkably, this magnetic transition is precisely coincident with a topological transition, characterized by a quantized jump in the Zak phase from $\pm\pi$ (topological) to $0$ (trivial). We systematically map out the phase boundaries in the parameter space spanned by the on-site energy staggering ($\epsilon$) and the inter-chain hopping asymmetry ($\lambda$). The resulting phase diagrams reveal the critical interaction strength required to induce the transition, providing a comprehensive and predictive picture of the system's ground state. Furthermore, by analyzing the energy spectrum, we distinguish the true ground state from metastable, quasi-stationary states that appear beyond the critical point. \textcolor{black}{In addition, we extend our analysis to the quantum geometry of the Bloch states by calculating the Fubini-Study metric \cite{Provost1980_FSMetric}. We show that the components of this metric tensor exhibit divergences that precisely coincide with the topological phase transitions. This provides a geometric interpretation of the transition, where the formation of Dirac cones in the band structure is signaled by singularities in the geometry of the quantum state space, a finding consistent with the results for the non-interacting case \cite{EspinosaChampo2023_Fubini}.}

The ability to control both magnetic and topological properties via a single interaction parameter suggests that correlated topological chains like the Creutz-Hubbard ladder could be promising platforms for future spintronic devices and quantum information processing. Our findings provide a detailed theoretical framework for understanding and predicting the behavior of such strongly correlated topological systems, paving the way for the exploration of interaction-induced topological phenomena in more complex materials.

The remainder of this paper is organized as follows. In Section \ref{sec:Hamiltonian Model}, we introduce the Creutz-Hubbard Hamiltonian and outline the self-consistent mean-field approximation used to solve it. In Section \ref{sec:results}, we present our main results, including the complete phase diagrams that map the topological and magnetic transitions as a function of the system's parameters, supported by an analysis of the energy spectrum and band structures. Finally, in Section \ref{sec:conclusions}, we summarize our conclusions and discuss the implications of our findings.

\section{Hamiltonian Model} \label{sec:Hamiltonian Model}

\begin{figure}
    \centering
    \includegraphics[width=1.0\linewidth]{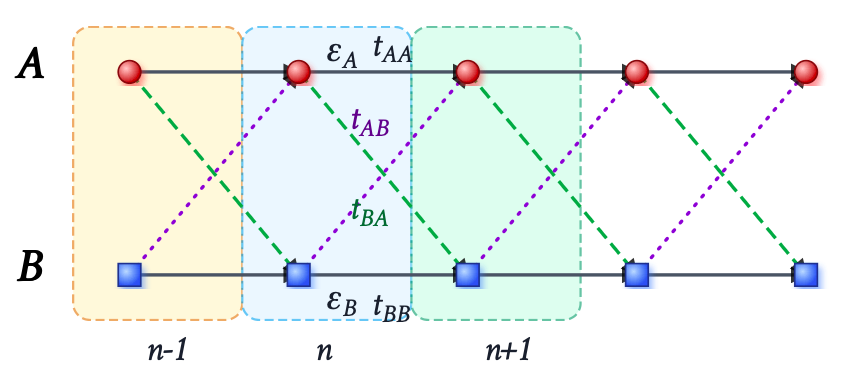}
    \caption{Schematic of the one-dimensional bipartite Creutz ladder. Red circles and blue squares represent sites on sublattices A and B, respectively.}
    \label{fig:model-creutz}
\end{figure}

Figure~\ref{fig:model-creutz} shows a one-dimensional bipartite Creutz ladder, where we consider interactions between spins located on each $A$ (red) and $B$ (blue) site. The one-band tight-binding Hubbard Hamiltonian for this system can be written as

\begin{equation}
    \label{eq:hamiltonian-hubbard-creutz}
    \begin{split}
    \mathcal{H}&=- \sum_{i, \sigma=\uparrow,\downarrow} \varepsilon_{i} c_{i,\sigma}^{\dagger}c_{i,\sigma}- \sum_{ij,\sigma=\uparrow,\downarrow} t_{ij} c^{\dagger}_{i, \sigma} c_{j,\sigma}+h.c\\
    &+ U \sum_{i} c^{\dagger}_{i, \uparrow} c^{\dagger}_{i,\downarrow}c_{i,\uparrow}c_{i,\downarrow} 
    \end{split}
\end{equation}
where $i,j$ are lattice site label, $\sigma$ a spin index, $\epsilon_i$ is the on-site energy, $t_ij$ the tranfer integral and $U$ the electron-electron interaction energy. $c_i$ is the anhilitation operator at site $i$. We now introduce the creation (annihilation) operators $\hat{a}_{n,\sigma}^{\dagger}(\hat{a}_{n,\sigma})$ and $\hat{b}_{n,\sigma}^{\dagger}(\hat{b}_{n,\sigma})$ acting on spin $\sigma$ at $A$ and $B$ sites of the $n$-th unit cell, respectively. The Hamiltonian can thus be rewritten as:

\begin{equation}
    \label{eq:hamiltonian-a-b-operators}
    \begin{split}
    \mathcal{H}&=H_{0}+H_{I},\\
    H_0&=- \sum_{n, \sigma=\uparrow,\downarrow} [\varepsilon_{A} \hat{a}^{\dagger}_{n,\sigma}\hat{a}_{n,\sigma}+\varepsilon_{B} \hat{b}^{\dagger}_{n,\sigma}\hat{b}_{n,\sigma}]\\
    & - \sum_{n, \sigma=\uparrow,\downarrow} [t_{AA} \hat{a}^{\dagger}_{n+1,\sigma}\hat{a}_{n,\sigma}+t_{BB} \hat{b}^{\dagger}_{n+1,\sigma}\hat{b}_{n,\sigma}]\\
    & - \sum_{n, \sigma=\uparrow,\downarrow} [t_{AB} \hat{a}^{\dagger}_{n+1,\sigma}\hat{b}_{n,\sigma}+t_{BA} \hat{b}^{\dagger}_{n+1,\sigma}\hat{a}_{n,\sigma}]+h.c.\\
   H_{I} &= U \sum_{n}[\hat{a}_{n,\uparrow}^{\dagger}\hat{a}_{n,\downarrow}^{\dagger} \hat{a}_{n, \uparrow} \hat{a}_{n,\downarrow}+\hat{b}_{n,\uparrow}^{\dagger}\hat{b}_{n,\downarrow}^{\dagger} \hat{b}_{n, \uparrow} \hat{b}_{n,\downarrow} ]
    \end{split}
\end{equation}

We rewrite $H_0$ using the Fourier transforms:
\begin{equation}
    \label{eq:opertors-k-space}
    \begin{split}
        \hat{a}_{k,\sigma}&= \frac{1}{\sqrt{N}} \sum_{n} \hat{a}_{n,\sigma} e^{i k x_{n}}\\
        \hat{b}_{k,\sigma}&= \frac{1}{\sqrt{N}} \sum_{n} \hat{b}_{n,\sigma} e^{i k x_{n}}\\
        \hat{a}_{k,\sigma}^{\dagger}&= \frac{1}{\sqrt{N}} \sum_{n} \hat{a}^{\dagger}_{n,\sigma} e^{-i k x_{n}}\\
        \hat{b}^{\dagger}_{k,\sigma}&= \frac{1}{\sqrt{N}} \sum_{n} \hat{b}_{n,\sigma}^{\dagger} e^{-i k x_{n}}\\
    \end{split}
\end{equation}
with $x_n = nl$, and $l$ the intercell spacing. Then,
\begin{equation}
    \label{eq:hamiltonian-a-b-operators-k}
    \begin{split}
    H_0(k)&=- \sum_{k, \sigma=\uparrow,\downarrow} [\varepsilon_{A} \hat{a}^{\dagger}_{k,\sigma}\hat{a}_{k,\sigma}+\varepsilon_{B} \hat{b}^{\dagger}_{k,\sigma}\hat{b}_{k,\sigma}]\\
    & - \sum_{k, \sigma=\uparrow,\downarrow} [t_{AA} \hat{a}^{\dagger}_{k,\sigma}\hat{a}_{k,\sigma}+t_{BB} \hat{b}^{\dagger}_{k,\sigma}\hat{b}_{k,\sigma}]e^{ikl}\\
    & - \sum_{k, \sigma=\uparrow,\downarrow} [t_{AB} \hat{a}^{\dagger}_{k,\sigma}\hat{b}_{k,\sigma}+t_{BA} \hat{b}^{\dagger}_{k,\sigma}\hat{a}_{k,\sigma}]e^{ikl}+h.c.
    \end{split}
\end{equation}

Following Ref.~\cite{Espinosa-Champo_2024}, we impose the constraints:
\begin{equation}
    \label{eq:conditions-previous-work}
    \begin{split}
        \varepsilon_A&=- \varepsilon_B \equiv \varepsilon\\
        t_{AA}&=-t_{BB} \equiv t_0\\
        t_{AB}&=-t_{BA}\equiv t_1
    \end{split}
\end{equation}
yielding
\begin{equation}
    \label{eq:hamiltonian-k-space}
    H_0(k)= - \sum_{k} \Psi_{k}^{\dagger}\mathcal{H}_{0}(k)\Psi_{k}, 
\end{equation}
where,
\begin{equation}
    \Psi_{k}=(\hat{a}_{k,\uparrow},\hat{b}_{k,\uparrow},\hat{a}_{k,\downarrow},\hat{b}_{k,\downarrow})^{T}.
\end{equation}
Here $ \mathcal{H}_{0}(k)$ is a $4 \times 4$ matrix defined as,
\begin{equation}
    \label{eq:hamiltonian-kitaev-like}
    \begin{split}
        \mathcal{H}_{0}(k)&= \mathbb{1}_{2\times 2} \otimes H_{Kit}(k) 
    \end{split}
\end{equation}
where $H_{Kit}(k)$ takes the exact form of the well known Kitaev Hamiltonian \cite{Leumer_2020},
\begin{equation}
     H_{Kit}(k)=[\epsilon+2t_0 \cos(kl)]\tau_{3}+2t_1 \sin(kl) \tau_{2}
\end{equation}
and $\tau_{i}, i=0,1,2,3$ is the $i$-th Pauli matrix. 

To treat the interaction term $H_I$ of the Hamiltonian, we adopt a mean-field approximation. Expressing the spin number operators for each sublattice $\alpha = A,B$ and spin $\sigma$ in the $j$-th cell as:
\begin{equation}
    \label{eq:spin-number-mean-field}
    \hat{n}_{\alpha,j,\sigma}=\langle \hat{n}_{\alpha,j,\sigma} \rangle +\Delta \hat{n}_{\alpha,j,\sigma}
\end{equation}

The interaction Hamiltonian is then approximated as:
\begin{equation}
    \label{eq:interaction-hamiltonian}
    \begin{split}
        H_{I}& \approx U \sum_{j, \alpha=A,B}\sum_{\sigma_1\neq \sigma_2}  \left[\langle \hat{n}_{\alpha,j,\sigma_{1}} \rangle \hat{n}_{\alpha,j,\sigma_{2}} -  \langle \hat{n}_{\alpha,j,\sigma_{1}} \rangle \langle \hat{n}_{\alpha,j,\sigma_{2}} \rangle \right]
    \end{split}
\end{equation}

Assuming a homogeneous system, the mean densities are independent of $j$:
\begin{equation}
    \label{eq:homogeneous-system}
    \langle \hat{n}_{\alpha,j,\sigma} \rangle\equiv \overline{n}_{\alpha,\sigma}
\end{equation}

Thus,
\begin{equation}
    \label{eq:interaction-hamiltonian-1}
    \begin{split}
        H_I&\approx H_{I}^{(0)}+H_{I}^{(1)},\\
        H_{I}^{(0)}&=-U  \sum_{\alpha=a,b; \sigma_1=\uparrow, \downarrow}\sum_{j, \sigma_2\neq \sigma_1} \overline{n}_{\alpha,\sigma_1}\overline{n}_{\alpha,\sigma_2},\\
        H_{I}^{(1)}&= U  \sum_{\alpha=a,b; \sigma_1=\uparrow, \downarrow}\sum_{j, \sigma_2\neq \sigma_1} \overline{n}_{\alpha,\sigma_1} \hat{n}_{\alpha,j,\sigma_2}
    \end{split}
\end{equation}

Since $H_{I}^{(0)}$ is a constant energy shift, only $H_{I}^{(1)}$ contributes to dynamics. Using Eq.~\eqref{eq:opertors-k-space}, we have:
\begin{equation}
    \label{eq:interaction-hamiltonian-2}
    \begin{split}
        H_{I}^{(1)}&= \sum_{k} \Psi_{k}^{\dagger} H_{I}^{(1)}(k) \Psi_{k}\\
        H_{I}^{(1)}(k)&=U\left( \begin{array}{cccc}
            \overline{n}_{a,\downarrow} & 0 & 0 & 0  \\
            0 & \overline{n}_{b,\downarrow} & 0 & 0\\ 
            0 & 0 & \overline{n}_{a,\uparrow} & 0\\
            0 & 0 & 0 & \overline{n}_{b,\uparrow}
        \end{array}\right)\\
        &= \frac{U}{2} \left( \begin{array}{cc}
            (\overline{n}_{a,\downarrow}+\overline{n}_{b,\downarrow}) \tau_0 &  0_{2\times2}\\
            0_{2\times 2} & (\overline{n}_{a,\uparrow}+\overline{n}_{b,\uparrow}) \tau_0 
        \end{array}\right)\\
        &+ \frac{U}{2} \left( \begin{array}{cc}
            (\overline{n}_{a,\downarrow}-\overline{n}_{b,\downarrow})\tau_3 &  0_{2\times2}\\
            0_{2\times 2} & (\overline{n}_{a,\uparrow}-\overline{n}_{b,\uparrow}) \tau_3
        \end{array}\right)
    \end{split}
\end{equation}
Therefore, the effective hamiltonian is
\begin{equation}
    \label{eq:effective-hamiltonian}
    H(k)= H_0(k)+H^{(1)}_{I}(k).
\end{equation}

\section{Results \label{sec:results}}

In this section, we present the results obtained analytically in a general manner and discuss several cases without considering whether the system could be stable, and then we present the results obtained numerically by applying SCF in the mean-field regime.

\subsection{Analytical results. \label{subsec:Analytical-results}}

As discussed in \cite{Espinosa-Champo_2024}, it is possible to reparameterize the on-site energy values $\varepsilon$, the intercell hopping parameter $t_1$, and the electronic interaction parameter $U$ in terms of the parameter $t_0$, as
\begin{equation}
    \label{eq:reparametrization-kitaev}
    \begin{split}
        \varepsilon&=\overline{\varepsilon} \, t_0,\\
        t_1&= \lambda \, t_0,\\
        U&= \overline{u} \, t_0,
    \end{split}
\end{equation}
so that, from Eq.~\eqref{eq:effective-hamiltonian}, the Hamiltonian can be rewritten in block form as
\begin{equation}
    \label{eq:hamiltonian-blocks}
    \begin{split}
        H(k)&= t_0 \begin{pmatrix}
            H_{\downarrow}(k) & 0_{2\times 2} \\
            0_{2 \times 2} &  H_{\uparrow}(k) 
        \end{pmatrix},
    \end{split}
\end{equation}
where
\begin{equation}
    \label{eq:hamiltonian-spin-blocks}
    H_{\sigma}(k)= \chi_{\sigma,+} \, \tau_{0} + (\chi_{\sigma,-}+\overline{\varepsilon}+2\cos(kl)) \, \tau_{3} + 2\lambda \, \sin(kl) \, \tau_{2},
\end{equation}
with $\chi_{\sigma,\pm}= \overline{u}(\overline{n}_{a,\sigma}\pm\overline{n}_{b,\sigma})/2$.

In this way, we can obtain the band expressions for each spin, given by
\begin{equation}
    \label{eq:bands-per-spin}
    \begin{split}
        E_{\eta,\sigma}(k)&= t_0 \Bigl\{ \chi_{\sigma,+}+\eta\bigl[(\chi_{\sigma,-}+\overline{\varepsilon}+2 \cos(kl))^{2} \\
        &\quad + (2\lambda \sin(kl))^{2}\bigr]^{1/2}\Bigr\},
    \end{split}
\end{equation}
where $\eta=\pm 1$.

From Eq.~\eqref{eq:bands-per-spin}, we define the normalized energies as
\begin{equation}
    \label{eq:bands-per-spin-renormalized}
    \begin{split}
\mathcal{E}_{\eta,\sigma}(k)\equiv \frac{E_{\eta,\sigma}(k)}{t_0}&=  \chi_{\sigma,+}+\eta\bigl[(\chi_{\sigma,-}+\overline{\varepsilon}+2 \cos(kl))^{2} \\
&\quad + (2\lambda \sin(kl))^{2}\bigr]^{1/2}.
\end{split}
\end{equation}

\begin{figure}[t]
    \centering
    \large{a)}\includegraphics[width=0.8\linewidth]{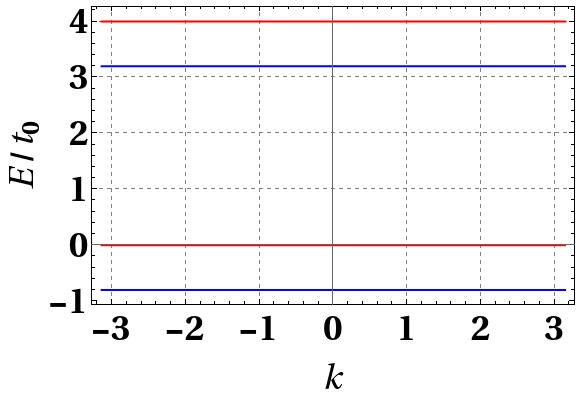}\\
    \large{b)}\includegraphics[width=0.8\linewidth]{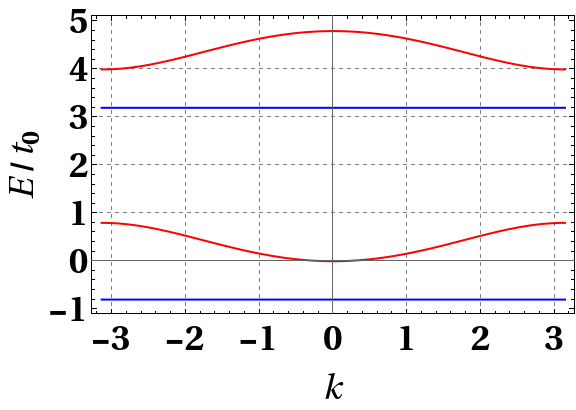}\\
    \large{c)}\includegraphics[width=0.8\linewidth]{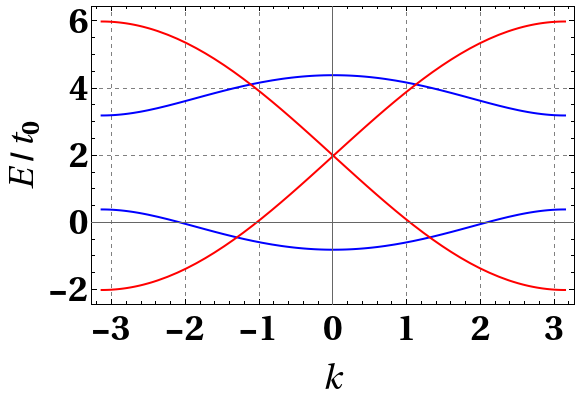}
    \caption{Band structure for the values $\overline{\varepsilon}=0,\lambda=1$ and $\overline{u}=4.0$. The blue and red lines correspond to spins $\downarrow$ and $\uparrow$, respectively. (a) For $\overline{n}_{a,\uparrow}=\overline{n}_{b,\uparrow}=0.5$ and $\overline{n}_{a,\downarrow}=\overline{n}_{b,\downarrow}=0.3$, non-degenerate flat bands arise because $\chi_{\sigma,-}=0$ (see Eq.~\eqref{eq:flat-bands-spin}). (b) Introducing an imbalance only in one spin, i.e., $\overline{n}_{a,\uparrow}=0.7,\overline{n}_{b,\uparrow}=0.5$ and $\overline{n}_{a,\downarrow}=\overline{n}_{b,\downarrow}=0.3$, yields dispersive bands for spin $\uparrow$ while spin $\downarrow$ remains flat. (c) As discussed in Eq.~\eqref{eq:eigenvalues-at-k-zero}, for $\overline{n}_{a,\uparrow}=0.0,\overline{n}_{b,\uparrow}=1.0$ and $\overline{n}_{a,\downarrow}=0.6,\overline{n}_{b,\downarrow}=0.3$, the condition $\chi_{\uparrow,-}=-2$ is satisfied, forming a Dirac cone for spin $\uparrow$ while opening a gap for spin $\downarrow$, rendering it an insulator; hence the system exhibits half-metallic behavior.}
    \label{fig:bands}
\end{figure}

\subsubsection{Electronic properties \label{subsub:electronic properties}}
From Eq.~\eqref{eq:bands-per-spin-renormalized}, the flat-band condition in the interacting case is obtained when $|\lambda|=1$ and $\overline{\varepsilon}+\chi_{\sigma,-}=0$, yielding flat bands with energy
\begin{equation}
    \label{eq:flat-bands-spin}
    \mathcal{E}_{\eta,\sigma}(k)= \chi_{\sigma,+}+2\eta.
\end{equation}
In Fig.~\ref{fig:bands}(a), the flat-band case analyzed in \cite{Espinosa-Champo_2024} is recovered for $\overline{u}=0$, and it is also possible to have flat bands for one spin and dispersive bands for the other as in Fig.~\ref{fig:bands}(b). Moreover, if $\overline{u}\neq 0$, the flat bands reported in \cite{Espinosa-Champo_2024} (i.e., $\overline{\varepsilon}=0,|\lambda|=1$) become dispersive.

On the other hand, in the long-wavelength limit, we can expand the spin Hamiltonian in Eq.~\eqref{eq:hamiltonian-spin-blocks} around $k=0$ to obtain
\begin{equation}
    \label{eq:hamiltonian-spin-block-at-k-zero}
    H_{\sigma}(k)\approx \chi_{\sigma,+} \, \tau_{0} + 2\lambda k l \, \tau_{2} + m \, \tau_{3}, \quad m=2+(\chi_{\sigma,-}+\overline{\varepsilon}),
\end{equation}
whose energy dispersion is
\begin{equation}
    \label{eq:eigenvalues-at-k-zero}
    \mathcal{E}_{\eta,\sigma}(k)\approx \chi_{\sigma,+}+\eta \sqrt{m^{2}+(2\lambda k l)^{2}}.
\end{equation}
Here, a gap $\Delta_{g}=2|m|$ closes when the critical value $\overline{\varepsilon}+\chi_{\sigma,-}=-2$ is reached, i.e., $m\approx0$, leading to linear Dirac-like dispersion. In Fig.~\ref{fig:bands}(c), since the spins are decoupled, one spin exhibits linear dispersion while the other remains insulating, thus displaying half-metal behavior.

\subsubsection{Topological properties \label{subsub:topological properties}}

From Eq.~\eqref{eq:hamiltonian-a-b-operators-k}, the Bloch states follow as
\begin{equation}
    \label{eq:eigenstates-hamiltonian-blochs}
    \begin{split}
        |\psi_{k,\downarrow}^{+}\rangle&= (\cos(\omega_{\downarrow}(k)/2), i\sin(\omega_{\downarrow}(k)/2),0,0)^{T},\\
        |\psi_{k,\downarrow}^{-}\rangle&= (i\sin(\omega_{\downarrow}(k)/2),\cos(\omega_{\downarrow}(k)/2),0,0)^{T},\\
        |\psi_{k,\uparrow}^{+}\rangle&= (0,0,\cos(\omega_{\uparrow}(k)/2), i\sin(\omega_{\uparrow}(k)/2))^{T},\\
        |\psi_{k,\uparrow}^{-}\rangle&= (0,0, i\sin(\omega_{\uparrow}(k)/2),\cos(\omega_{\uparrow}(k)/2))^{T},
    \end{split}
\end{equation}
where we define
\begin{equation}
    \label{eq:omega-spin}
    \begin{split}
    \omega_{\sigma}(k)&=\arg[(\chi_{\sigma,-}+\overline{\varepsilon}+2\cos(kl))+2i\lambda\sin(kl)]\\
    &=\arctan\Bigl(\frac{2\lambda\sin(kl)}{\Sigma_{\sigma}+2\cos(kl)}\Bigr),
    \end{split}
\end{equation}
with $\Sigma_{\sigma}\equiv \chi_{\sigma,-}+\overline{\varepsilon}$ and $\arg(z)$ denoting the principal argument of $z\in\mathbb{C}$.

\begin{figure}
    \centering
    \includegraphics[width=0.99\linewidth]{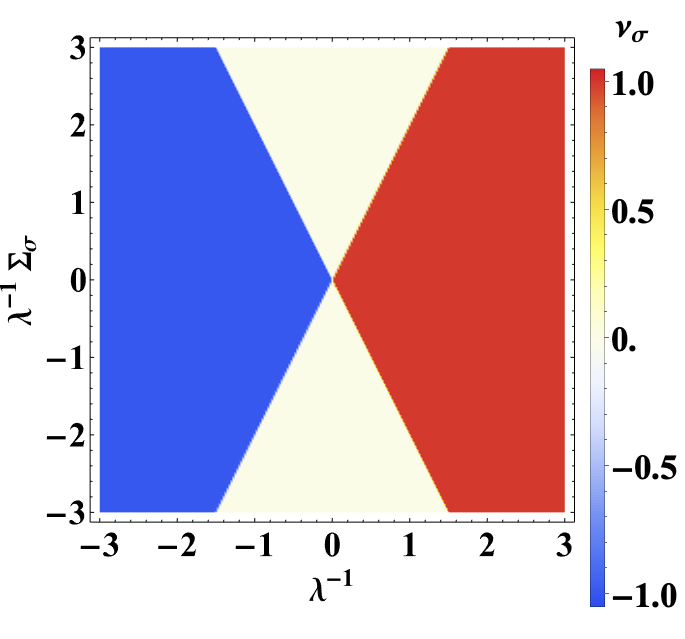}
    \caption{Topological phase diagram for each spin block Hamiltonian \eqref{eq:hamiltonian-spin-blocks}, calculated numerically and in agreement with Eq.~\eqref{eq:winding-spin}. The phase boundaries are determined by $|\Sigma_{\sigma}|=2$, indicating the emergence of linear (Dirac) dispersion in the band structure.}
    \label{fig:winding-number}
\end{figure}

The winding number for each spin is defined as \cite{Espinosa-Champo_2024}
\begin{equation}
    \label{eq:winding-spin}
    \begin{split}
    \nu_{\sigma}&= \frac{1}{2\pi} \int_{-\pi/l}^{\pi/l} dk\, \partial_{k} \omega_{\sigma}(k),\\
    &= \frac{1}{2\pi} \int_{-\pi/l}^{\pi/l} dk\, \frac{2\lambda\bigl(2+\Sigma_{\sigma}\cos(kl)\bigr)}{(\Sigma_{\sigma}+2\cos(kl))^{2}+4\lambda^{2}\sin^{2}(kl)}\\
    &= \begin{cases}
        1 & -2 \lambda^{-1} \leq \lambda^{-1}\Sigma_{\sigma} \leq 2\lambda^{-1},\\
        -1 & 2\lambda^{-1} \leq \lambda^{-1}\Sigma_{\sigma} \leq -2\lambda^{-1},\\
        0 & \text{otherwise},
    \end{cases}
    \end{split}
\end{equation}
where $\partial_{k}\omega_{\sigma}(k)$ is the winding-number density \cite{Espinosa-Champo_2024}. In Fig.~\ref{fig:winding-number}, the topological phase diagram for each spin block of the Hamiltonian \eqref{eq:hamiltonian-spin-blocks} is shown, demonstrating that one can artificially engineer a nontrivial topology for one spin while the other remains trivial.

\subsubsection{Fubini–Study metric}
  \begin{figure*}[t]
    \centering
    \large{a)}\includegraphics[width=0.45\linewidth]{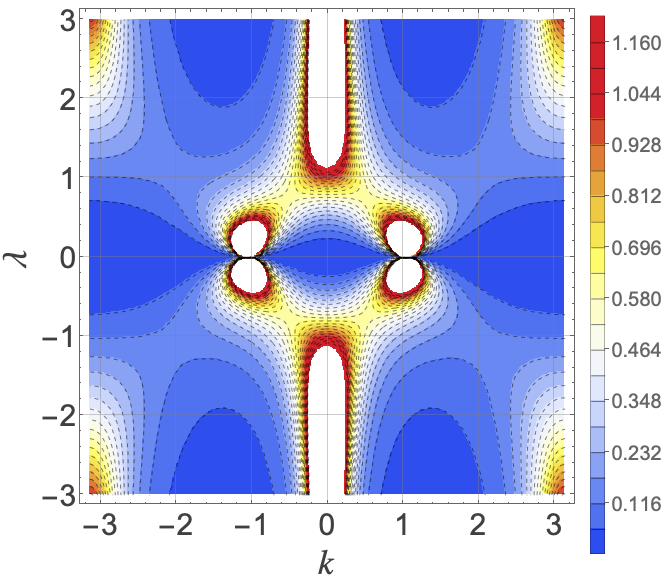}
    \large{b)}\includegraphics[width=0.45\linewidth]{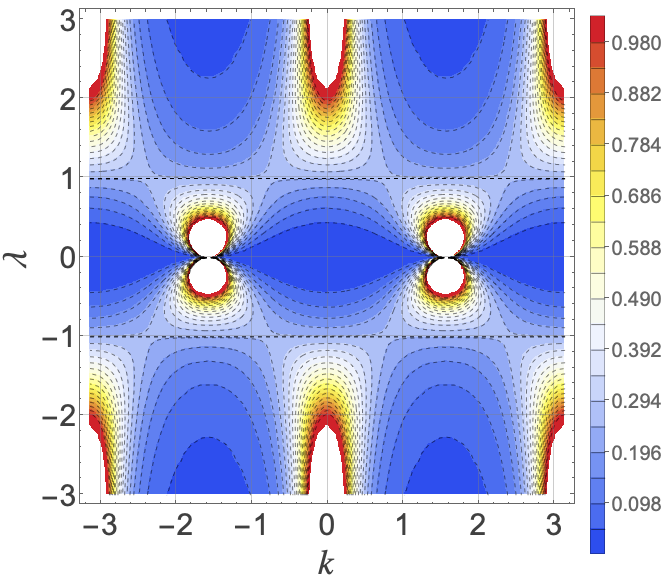}\\
    \large{c)}\includegraphics[width=0.45\linewidth]{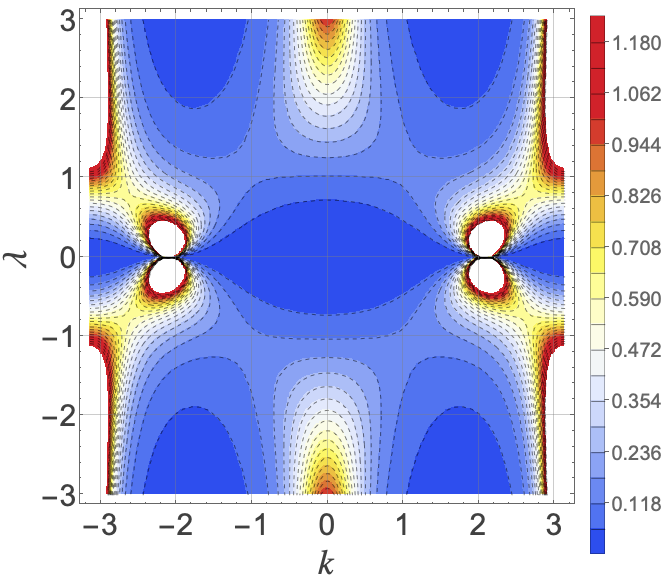}
    \caption{Contour plot of the Fubini–Study metric components $g_{kk,\sigma\sigma}^{\eta\eta}$ (Eq.~\eqref{eq:fubini-study-detailed}) as a function of $k$ and $\lambda$ for $\overline{\varepsilon}=0$, $l=1$, and (a) $\chi_{\sigma,-}=-1.0$, (b) $\chi_{\sigma,-}=0.0$, and (c) $\chi_{\sigma,-}=1.0$. In case (b), when the spin densities are equal ($\overline{n}_{a,\sigma}=\overline{n}_{b,\sigma}$), the system retains flat bands for $|\lambda|=1$, whereas in the other cases the geometry deforms due to spin-density imbalance. White regions indicate Dirac cones forming near $k=0$ or $k=\pm\pi/l$, as detailed further in the \textit{Supplementary Material A} .}
    \label{fig:quantum-metric-different-spin-densities}
\end{figure*}
The quantum geometric tensor is a fundamental quantity for understanding the behavior and characteristics of quantum systems. Recent studies have linked it to phenomena such as superfluidity \cite{verma2025}, quantum capacitance \cite{Komissarov2024}, and it is useful for analyzing topological insulators and flat-band materials. Moreover, it has gained prominence for fundamentally describing the emergence of electronic structure and its properties in materials \cite{yu2025}.

In general, consider a quantum state $|\phi(\boldsymbol{\xi})\rangle$ in an $N$-dimensional parameter space, where $\boldsymbol{\xi}=(\xi_{1},\ldots,\xi_{N})$ is a set of parameters. The geometry of this space is characterized by the quantum geometric tensor  \cite{TomokiOzawa2018, Aleksi2021, Cayssol_2021, Kruchkov2022, BernevigBogdan2022, TomokiOzawa2019}, given by
\begin{equation}
    \label{eq:quantum-metric-tensor-definition}
    Q_{\mu\nu}(\boldsymbol{\xi})= \langle\partial_{\mu}\phi(\boldsymbol{\xi})|\mathbb{P}_{\phi(\boldsymbol{\xi})}|\partial_{\nu}\phi(\boldsymbol{\xi})\rangle,
\end{equation}
where $\mathbb{P}_{\phi(\boldsymbol{\xi})}=\mathbb{1}-|\phi(\boldsymbol{\xi})\rangle\langle\phi(\boldsymbol{\xi})|$ is the orthogonal complement projector.

Since $Q_{\mu\nu}$ may be complex, its real part defines the Fubini–Study metric,
\begin{equation}
    \label{eq:fubini-study-metric}
    g_{\mu\nu}(\boldsymbol{\xi})= \mathrm{Re}[Q_{\mu\nu}(\boldsymbol{\xi})],
\end{equation}
and its imaginary part is associated with the Berry curvature $\Omega_{\mu\nu}(\boldsymbol{\xi})$,
\begin{equation}
    \label{eq:Berry-curvature}
    \mathrm{Im}[Q_{\mu\nu}(\boldsymbol{\xi})]= -\tfrac{1}{2}\Omega_{\mu\nu}(\boldsymbol{\xi}).
\end{equation}
The Fubini–Study metric measures the statistical distance between pure states $|\phi(\boldsymbol{\xi})\rangle$ and $|\phi(\boldsymbol{\xi}+d\boldsymbol{\xi})\rangle$, providing a way to distinguish them \cite{bengtsson_zyczkowski_2006}.

If we parameterize functions in momentum space $k$, we use the Bloch basis states $|\rho(k)\rangle$, and Eq.~\eqref{eq:fubini-study-metric} becomes
\begin{equation}
    \label{eq:fubini-study-k-space}
    g_{kk}= \langle\partial_{k}\rho(k)|\partial_{k}\rho(k)\rangle - \langle\partial_{k}\rho(k)|\rho(k)\rangle\langle\rho(k)|\partial_{k}\rho(k)\rangle.
\end{equation}
In our case, using the spin-dependent Bloch states in Eq.~\eqref{eq:eigenstates-hamiltonian-blochs}, we define the Fubini–Study metric for this system as
\begin{equation}
    \label{eq:fubini-study-creutz}
    g_{kk,\sigma\sigma'}^{\eta,\eta'}= \langle\partial_{k}\psi_{k,\sigma}^{\eta}|\partial_{k}\psi_{k,\sigma'}^{\eta'}\rangle - \langle\partial_{k}\psi_{k,\sigma}^{\eta}|\psi_{k,\sigma'}^{\eta'}\rangle\langle\psi_{k,\sigma'}^{\eta'}|\partial_{k}\psi_{k,\sigma}^{\eta}\rangle,
\end{equation}
with $\sigma,\sigma'=\uparrow,\downarrow$ and $\eta,\eta'=\pm$.

The derivatives of the eigenstates are
\begin{equation}
    \label{eq:derivatives-eigenstates-hamiltonian-blochs}
    \begin{split}
    |\partial_k \psi_{k,\downarrow}^{+}\rangle&=\tfrac{\partial_{k}\omega_{\downarrow}(k)}{2}(-\sin(\tfrac{\omega_{\downarrow}(k)}{2}), i\cos(\tfrac{\omega_{\downarrow}(k)}{2}),0,0)^{T},\\
        |\partial_k \psi_{k,\downarrow}^{-}\rangle&=\tfrac{\partial_{k}\omega_{\downarrow}(k)}{2}(i\cos(\tfrac{\omega_{\downarrow}(k)}{2}),-\sin(\tfrac{\omega_{\downarrow}(k)}{2}),0,0)^{T},\\
        |\partial_k \psi_{k,\uparrow}^{+}\rangle&=\tfrac{\partial_{k}\omega_{\uparrow}(k)}{2}(0,0,-\sin(\tfrac{\omega_{\uparrow}(k)}{2}), i\cos(\tfrac{\omega_{\uparrow}(k)}{2}))^{T},\\
        |\partial_k \psi_{k,\uparrow}^{-}\rangle&=(0,0,i\cos(\tfrac{\omega_{\uparrow}(k)}{2}),-\sin(\tfrac{\omega_{\uparrow}(k)}{2}))^{T}.
    \end{split}
\end{equation}

From Eqs.~\eqref{eq:fubini-study-creutz} and \eqref{eq:derivatives-eigenstates-hamiltonian-blochs}, we find
\begin{equation}
    \label{eq:fubini-study-detailed}
    g_{kk,\sigma\sigma'}^{\eta \eta'}= \begin{cases}
            0, & \sigma \neq \sigma'\\
            \eta\eta'\bigl(\tfrac{\partial_{k}\omega_{\sigma}(k)}{2}\bigr)^{2}, & \sigma=\sigma',
        \end{cases}
\end{equation}

The former case is expected since the spins are decoupled; for $\eta=-\eta'$ it arises from particle-hole symmetry. Similar relations to the spinless case are maintained \cite{EspinosaChampo2023_Fubini}, and using Eq.~\eqref{eq:winding-spin}, one obtains
\begin{equation}
    \label{eq:relations-like-creutz}
    \begin{split}
        \sqrt{|\det(g_{kk,\sigma\sigma}^{\eta \eta'})|}&= \tfrac{|\partial_{k}\omega_{\sigma}(k)|}{2}\\
        \pi |\nu_{\sigma}| \leq \mathrm{vol}(\mathrm{BZ}) &\leq \tfrac{1}{2} \mathrm{vol}(S^{1}),
    \end{split}
\end{equation}
where
\begin{equation}
    \label{eq:definitions-volumes}
    \begin{split}
        \mathrm{vol}(\mathrm{BZ})&\equiv \int_{\mathrm{BZ}} \sqrt{|g_{kk,\sigma\sigma}^{\eta \eta}|}\,dk,\\
        \mathrm{vol}(S^{1})&\equiv \int_{S^{1}} d\theta.
    \end{split}
\end{equation}
This implies that the geometry of flat bands can be described by the Fubini–Study metric through a conformal transformation, as discussed in Ref.~\cite{Espinosa-Champo_2024}.

In Fig.~\ref{fig:quantum-metric-different-spin-densities}, the contour plot of $g_{kk,\sigma\sigma}^{\eta\eta}$ vs. $k$ and $\lambda$ for $\overline{\varepsilon}=0$, $l=1$, and $\chi_{\sigma,-}=-1.0,0.0,1.0$ is shown. As observed, Dirac cone formation depends on the spin-density imbalance $\chi_{\sigma,-}=\overline{n}_{a,\sigma}-\overline{n}_{b,\sigma}$: for $\chi_{\sigma,-}\le0$ cones form near $k=0$, while for $\chi_{\sigma,-}\ge0$ they form near $k=\pm\pi/l$ (see also the Supplementary Material A). When the imbalance vanishes ($\chi_{\sigma,-}=0$), flat bands persist for that spin.

\subsection{Numerical results \label{subsec:numerical results}}

To validate our analytical predictions and explore the full parameter space, we performed extensive self-consistent mean-field calculations. These were implemented using the \texttt{sisl} library for efficient Hamiltonian construction \cite{zerounian2020sisl} and the \texttt{hubbard} package for the self-consistent loop \cite{sanzwuhl2023hubbard}. We systematically varied the dimensionless parameters for the on-site energy ($\bar{\epsilon}$), the inter-chain hopping asymmetry ($\lambda$), and the Hubbard interaction ($\bar{u}$). For each point in this parameter space, the mean-field densities $\bar{n}_{\alpha,\sigma}$ were iterated until convergence, allowing for the calculation of the system's ground state energy, the Zak phase, and the spatial spin polarization.

To characterize the topological nature of the system's ground state, we compute the Zak phase for each spin channel $\sigma$. The Zak phase is a Berry phase acquired by an electron as it moves across the one-dimensional Brillouin zone and serves as a bulk topological invariant for 1D systems \cite{Zak1989}. For a given spin and band $\eta$, it is defined as the integral of the Berry connection,
\begin{equation}
\label{eq:Zak_phase}
    \gamma_{\sigma}^{\eta} = i \int_{-\pi/a}^{\pi/a} \langle u_{k,\sigma}^{\eta} | \partial_k u_{k,\sigma}^{\eta} \rangle \, dk,
\end{equation}
where $|u_{k,\sigma}^{\eta}\rangle$ is the periodic part of the Bloch wave function. For systems with chiral symmetry, such as the Creutz ladder, the Zak phase is quantized to either $0$ or $\pi$ (modulo $2\pi$). This quantity is directly related to the winding number $\nu_\sigma$, a topological integer that counts how many times the Hamiltonian vector wraps around the origin in momentum space \cite{Asboth2016_Book}. The relationship is given by $\gamma_\sigma = \pi \nu_\sigma \pmod{2\pi}$. A non-zero winding number ($\nu_\sigma = \pm 1$, corresponding to a Zak phase of $\pm\pi$) indicates a topologically non-trivial phase in the context of chiral symmetric models, while a zero winding number ($\nu_\sigma = 0$, Zak phase of $0$) indicates a trivial phase.

Our central numerical finding is the existence of an interaction-driven, first-order phase transition that is simultaneously magnetic and topological. This phenomenon is best illustrated by examining the system's behavior for a representative set of parameters as a function of increasing interaction strength $\bar{u}$. 


\begin{figure}[h!]
    \centering
    a)\includegraphics[scale=0.3]{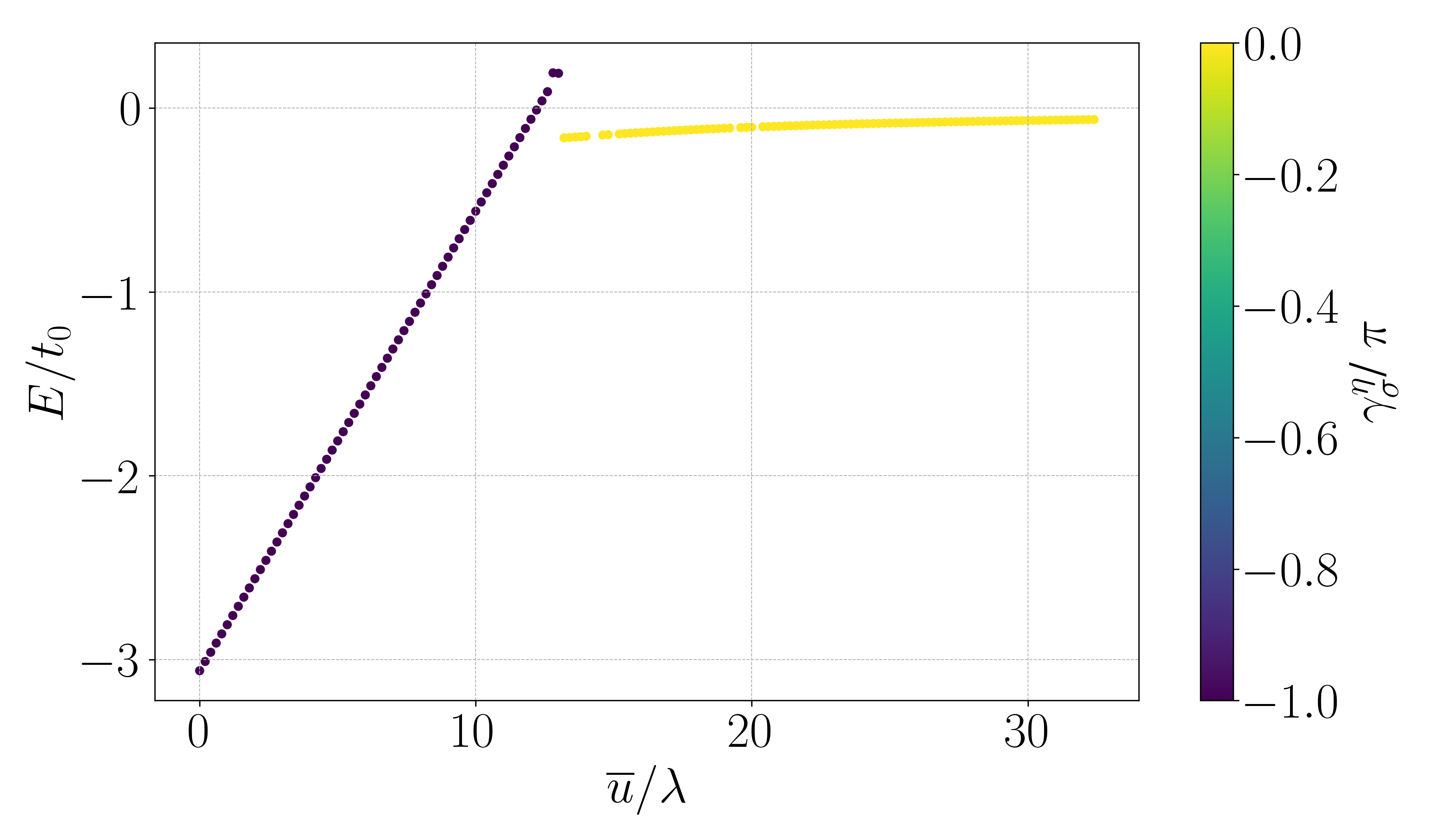}\\
    b)\includegraphics[scale=0.3]{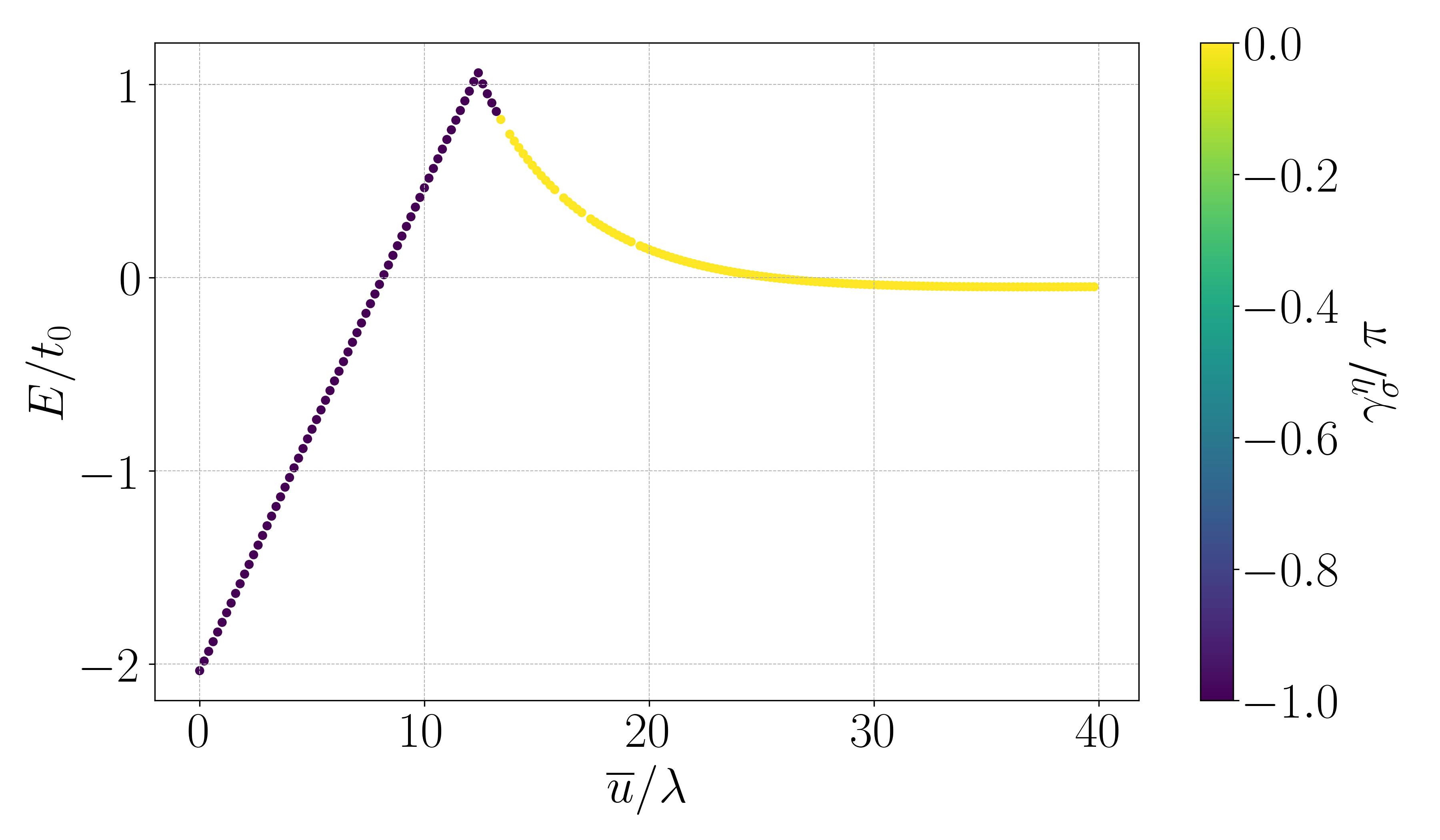}
    \caption{Total energy per site as a function of the normalized interaction $\overline{u}/\lambda$ for fixed $\bar{\epsilon}=0.0$ and $\lambda=0.5$, shown for two different temperatures: (a) low temperature ($kT=0.25$) and (b) high temperature ($kT=1.0$). The color of each point corresponds to the Zak phase (Eq. \ref{eq:Zak_phase}). At low temperature, a sharp change indicates a first-order transition. At higher temperature, the transition becomes smoother, suggesting thermal fluctuations smear the sharp boundary.}
    \label{fig:energy_bifurcation}
\end{figure}

As shown in Fig. \ref{fig:energy_bifurcation}, for small values of $\bar{u}$, the system's energy decreases linearly, and the Zak phase remains quantized at $\pm\pi$ (indicated by the dark color), corresponding to a topologically non-trivial, anti-ferromagnetic (AF) state. However, upon reaching a critical interaction strength $U_c$, the energy spectrum bifurcates. The system undergoes a first-order transition, jumping to a new, lower-energy ground state. This new ground state is characterized by a Zak phase quantized at $0$ (bright color), signaling its trivial topological nature. The upper energy branch that emerges beyond $U_c$ corresponds to a metastable solution of the self-consistent equations, not the true ground state.

This topological transition is perfectly coincident with a magnetic one. By analyzing the spatial distribution of the spin densities, we confirm that the system transitions from an anti-ferromagnetic (AF) configuration for $U < U_c$ to a ferromagnetic (F) one for $U > U_c$.


\begin{figure}[h!]
 a) \includegraphics[scale=0.25]{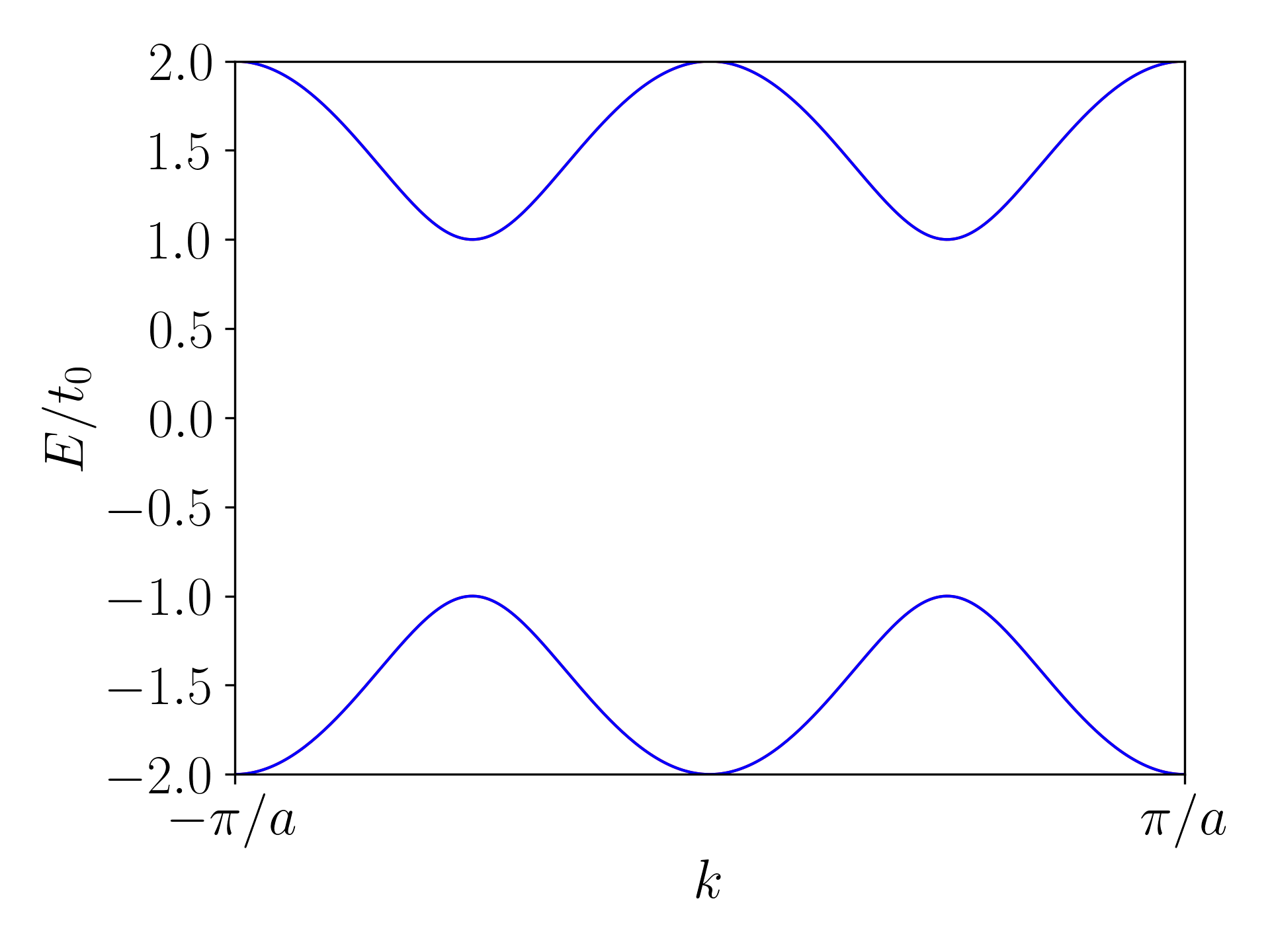}
 b) \includegraphics[scale=0.36]{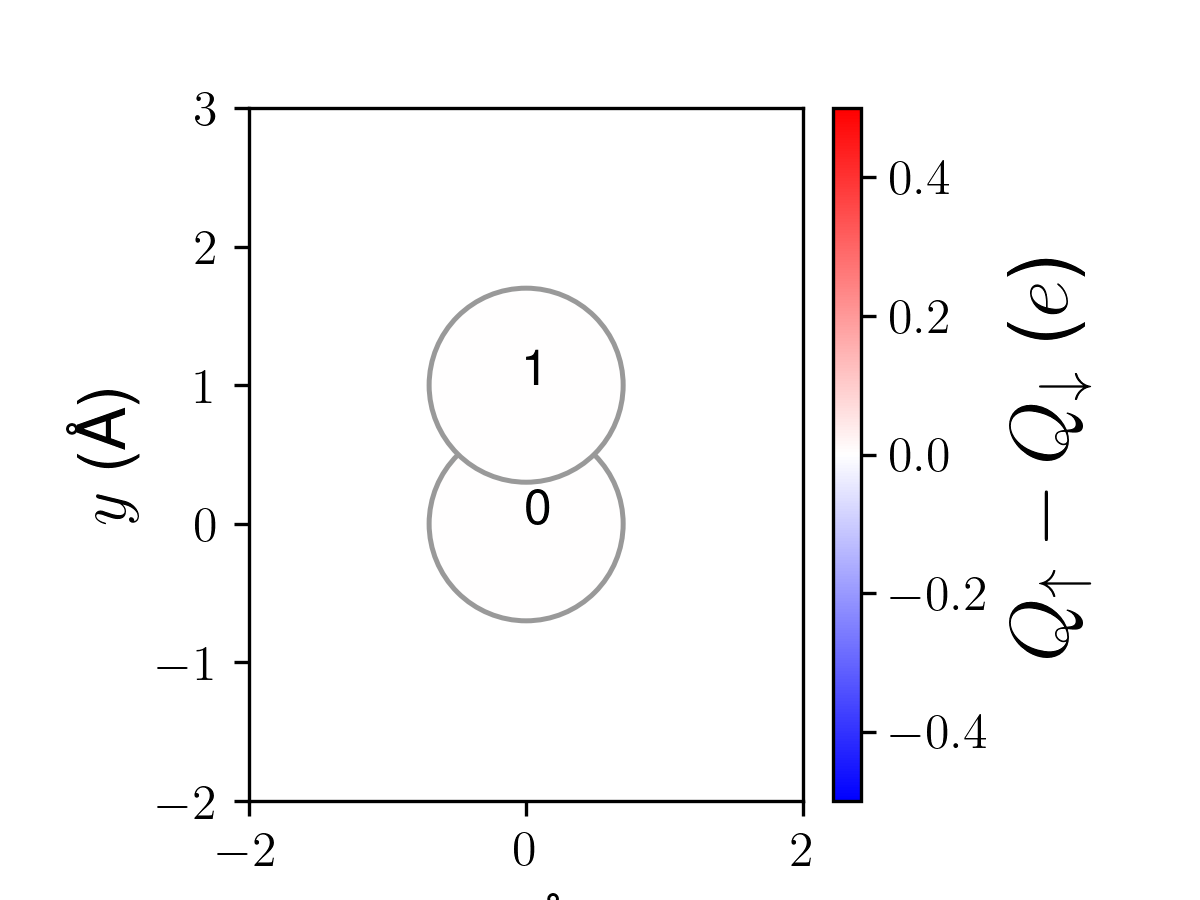}\\
 c) \includegraphics[scale=0.25]{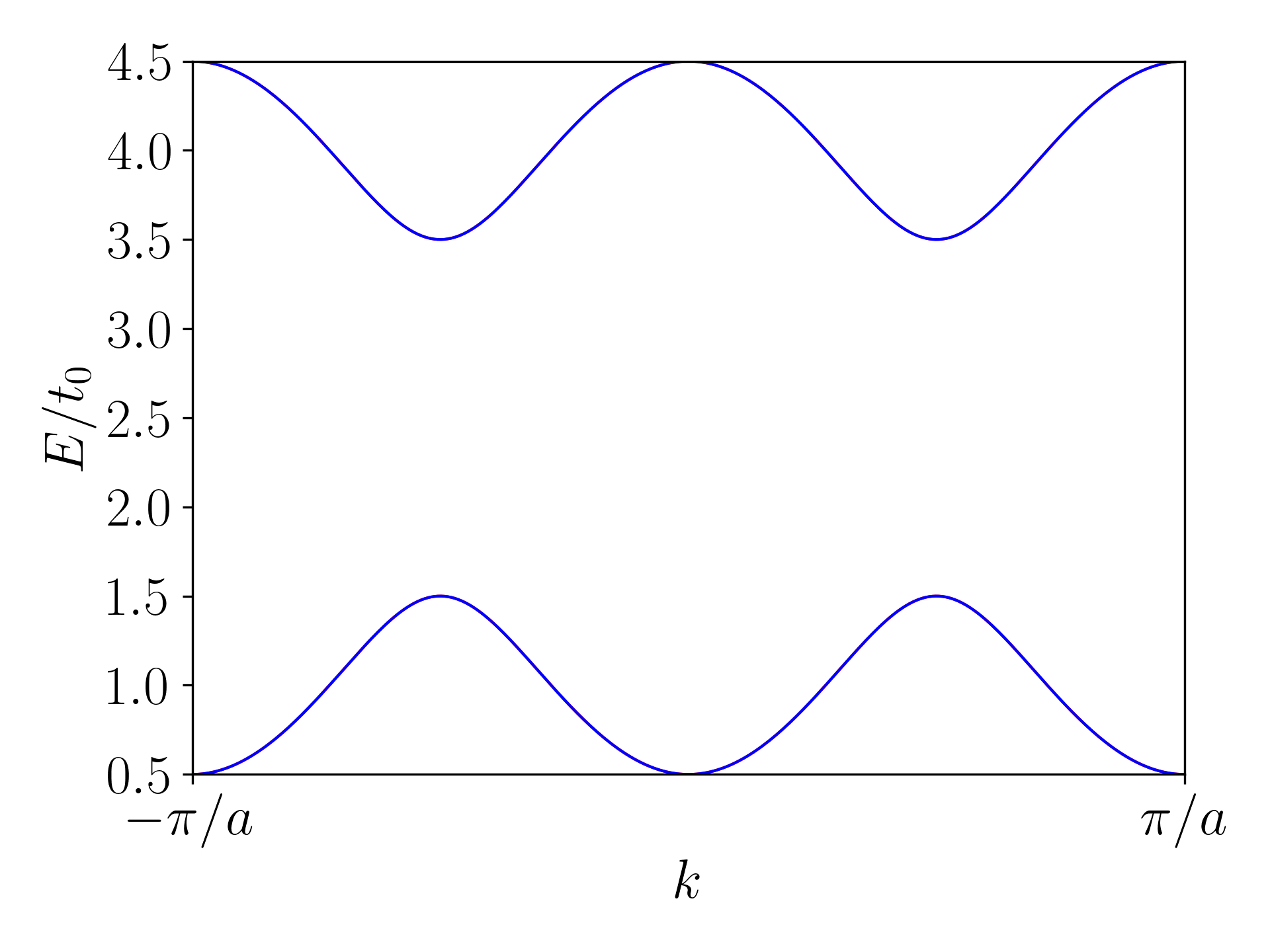}
 d) \includegraphics[scale=0.36]{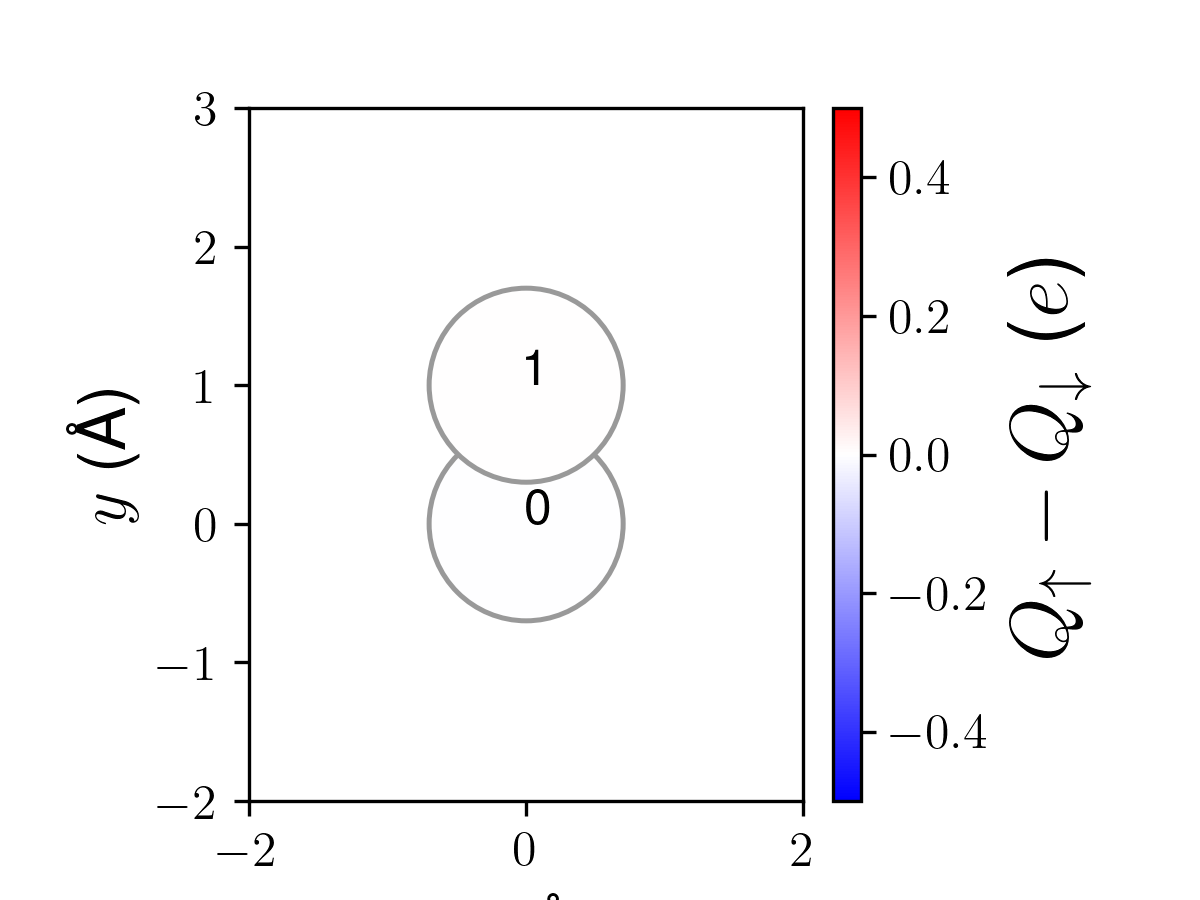}\\
 e) \includegraphics[scale=0.25]{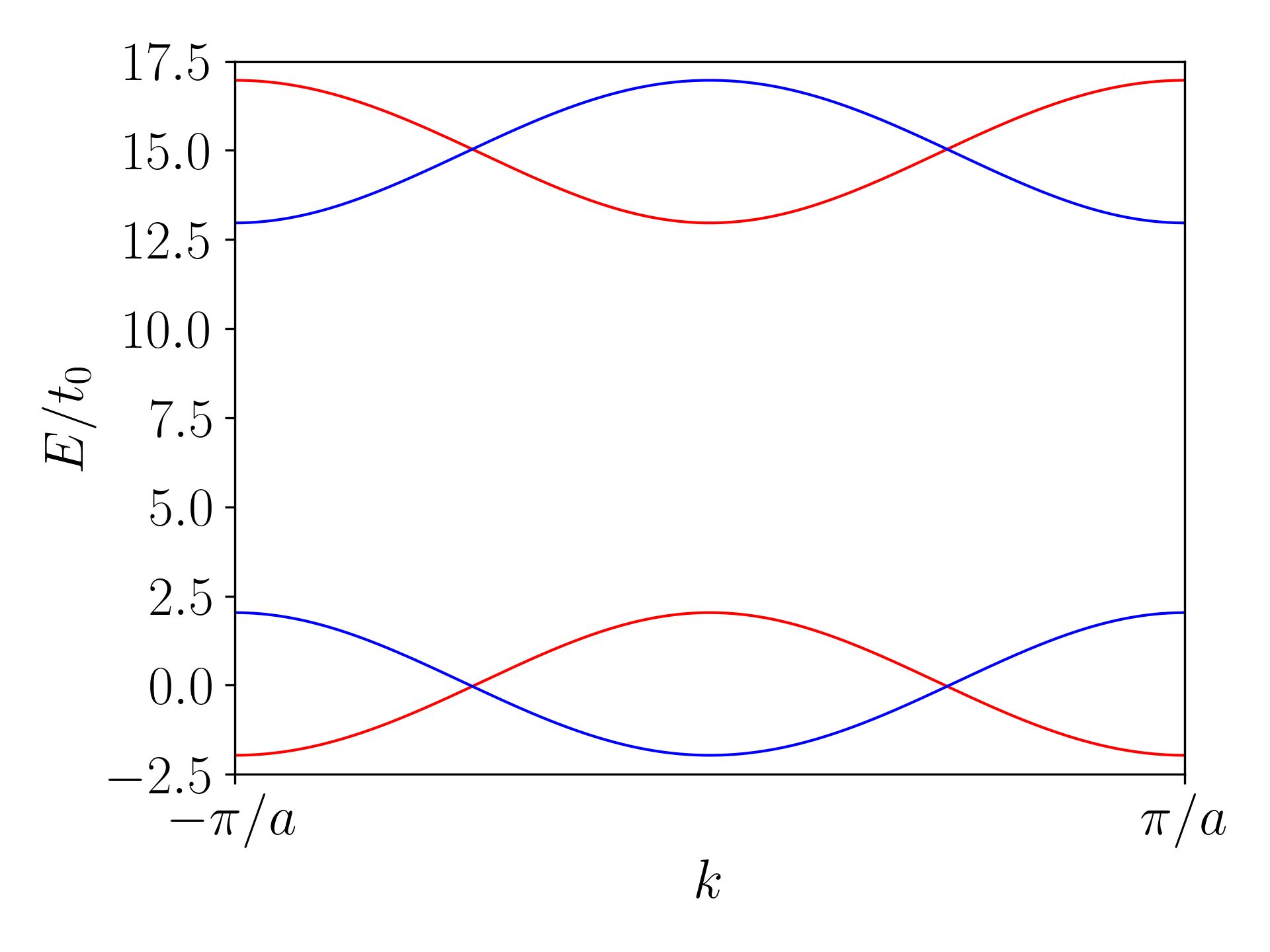}
 f) \includegraphics[scale=0.36]{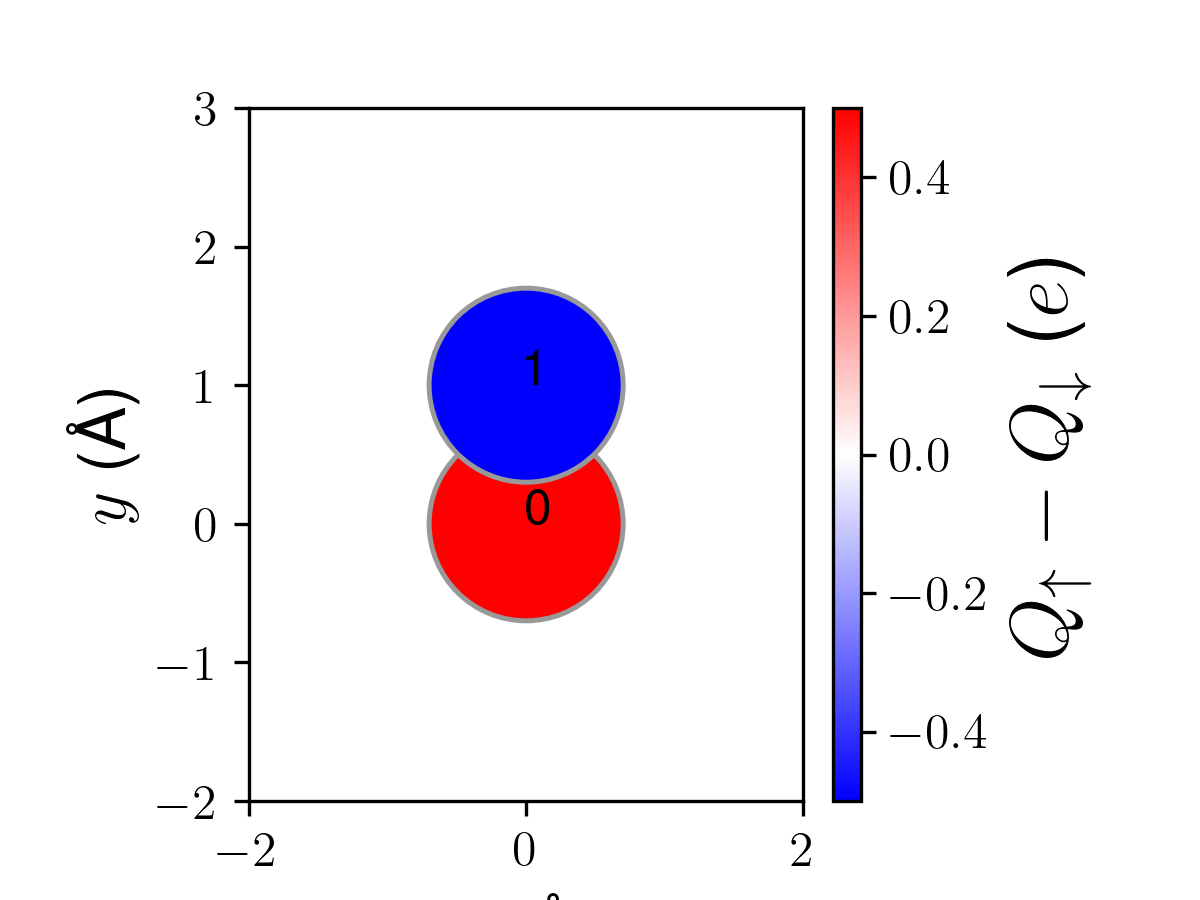}
    \caption{Band Structure and spin polarization on the two sites of a unit cell for fixed $\overline{\epsilon}=0.0$, $\lambda=0.5$ and (a-b) $\overline{u}=0$, (c-d) $\overline{u}=5.0$, i.e., $\overline{u} < \overline{u}_c$, showing an anti-ferromagnetic (AF) alignment, and (e-f) with $\overline{u}=15.0$,i.e., $\overline{u} > \overline{u}_c$, showing a ferromagnetic (F) alignment. This confirms the magnetic nature of the interaction-driven transition.}
    \label{fig:spin_polarization}
\end{figure}

\begin{figure}[h!]
    \centering
 a)\includegraphics[scale=0.18]{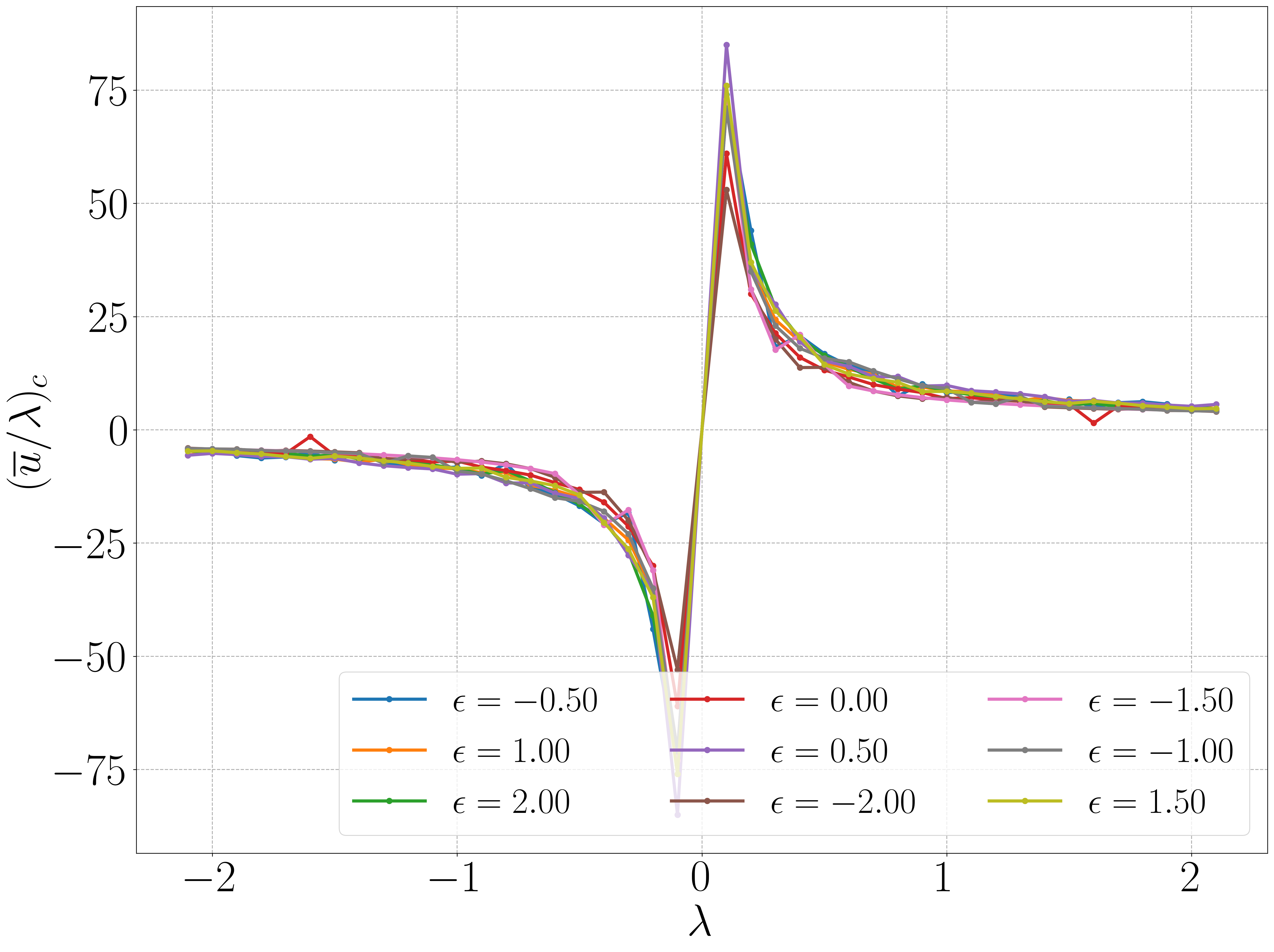}\\
  b)\includegraphics[scale=0.18]{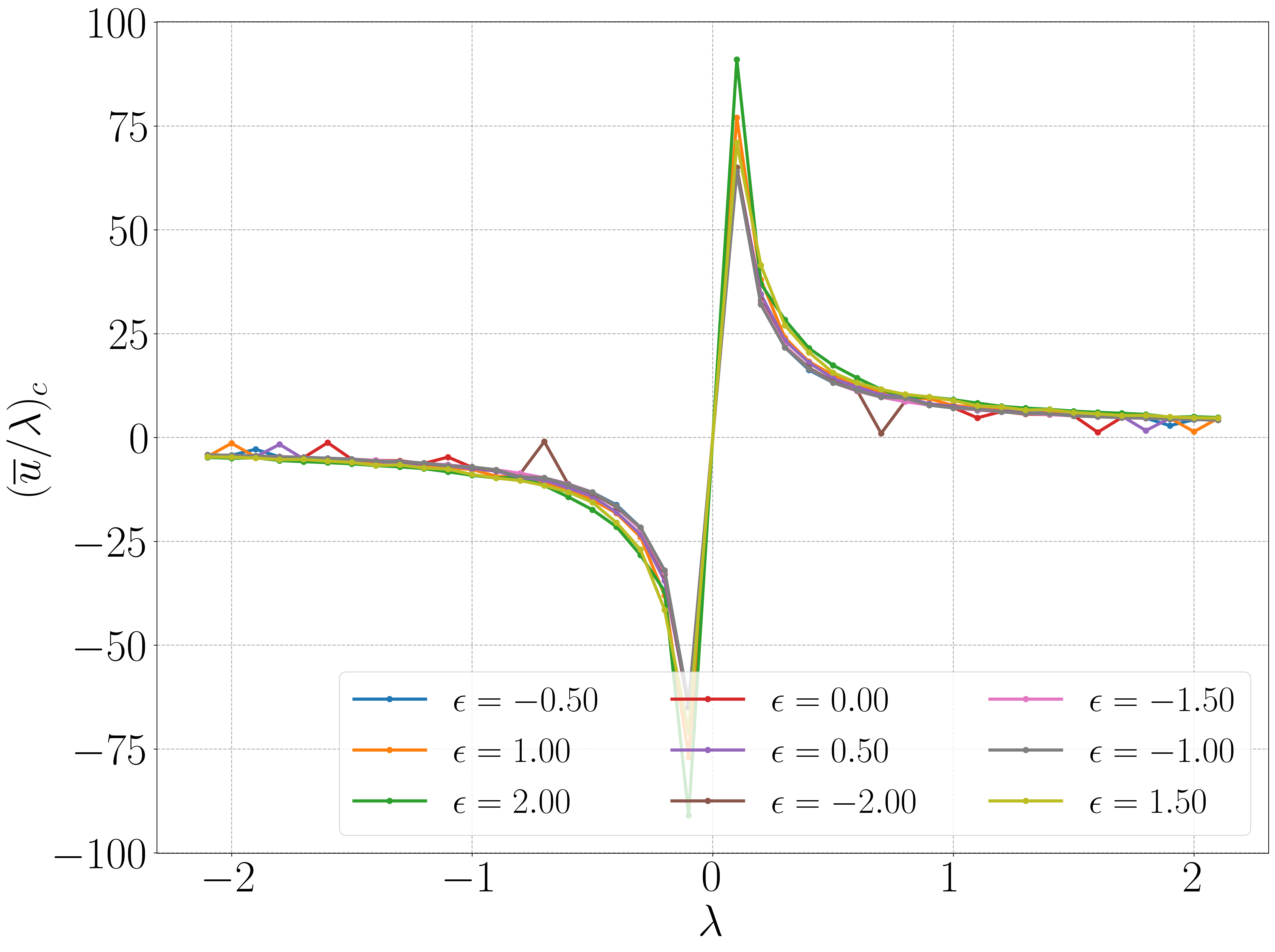}
    \caption{Combined phase diagram showing the critical bifurcation point $(\overline{u}/\lambda)_c$ as a function of $\lambda$ for several fixed values of $\bar{\epsilon}$. Panel (a) correspond to a low temperature ($kT=0.25$) eV, while panel (b) represent a higher temperature ($kT=1.0$) eV. An increase in temperature generally shifts the phase boundary, requiring a stronger interaction to induce the topological phase.}
    \label{fig:phase_diagrams}
\end{figure}

To systematically map the phase boundaries, we identified the critical interaction strength, $U_c$, for each pair of $(\bar{\epsilon}, \lambda)$ parameters, using the robust criterion of the Zak phase jump from $|\nu_\sigma| \approx 1$ to $\nu_\sigma = 0$. The results are summarized in the phase diagrams presented in Fig. \ref{fig:phase_diagrams}. These diagrams illustrate how the critical interaction required to induce the topological phase depends on the underlying system parameters and temperature.

\begin{figure*}[h!]
    \centering
 a) \includegraphics[scale=0.18]{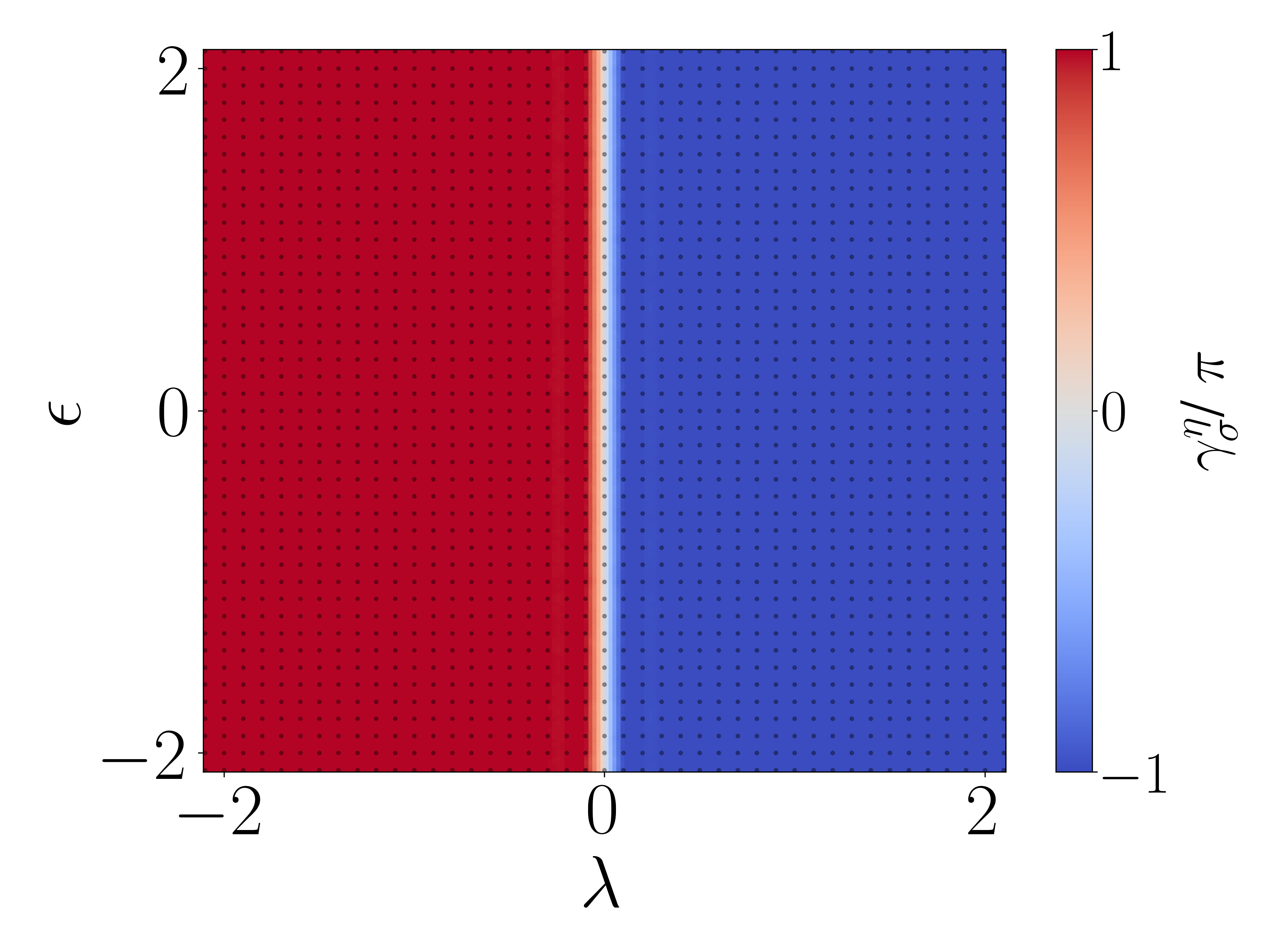}
 b) \includegraphics[scale=0.18]{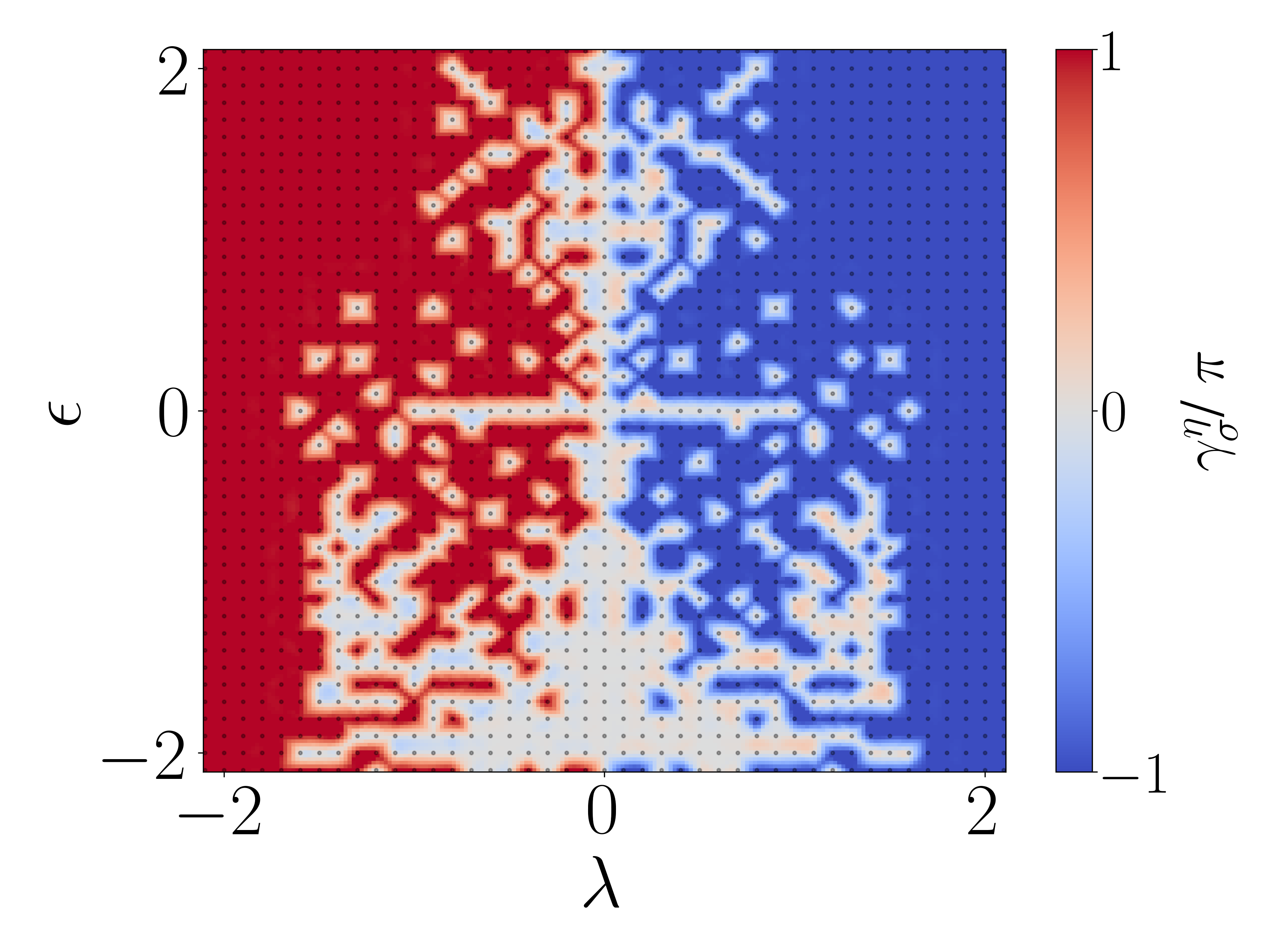}
 c) \includegraphics[scale=0.18]{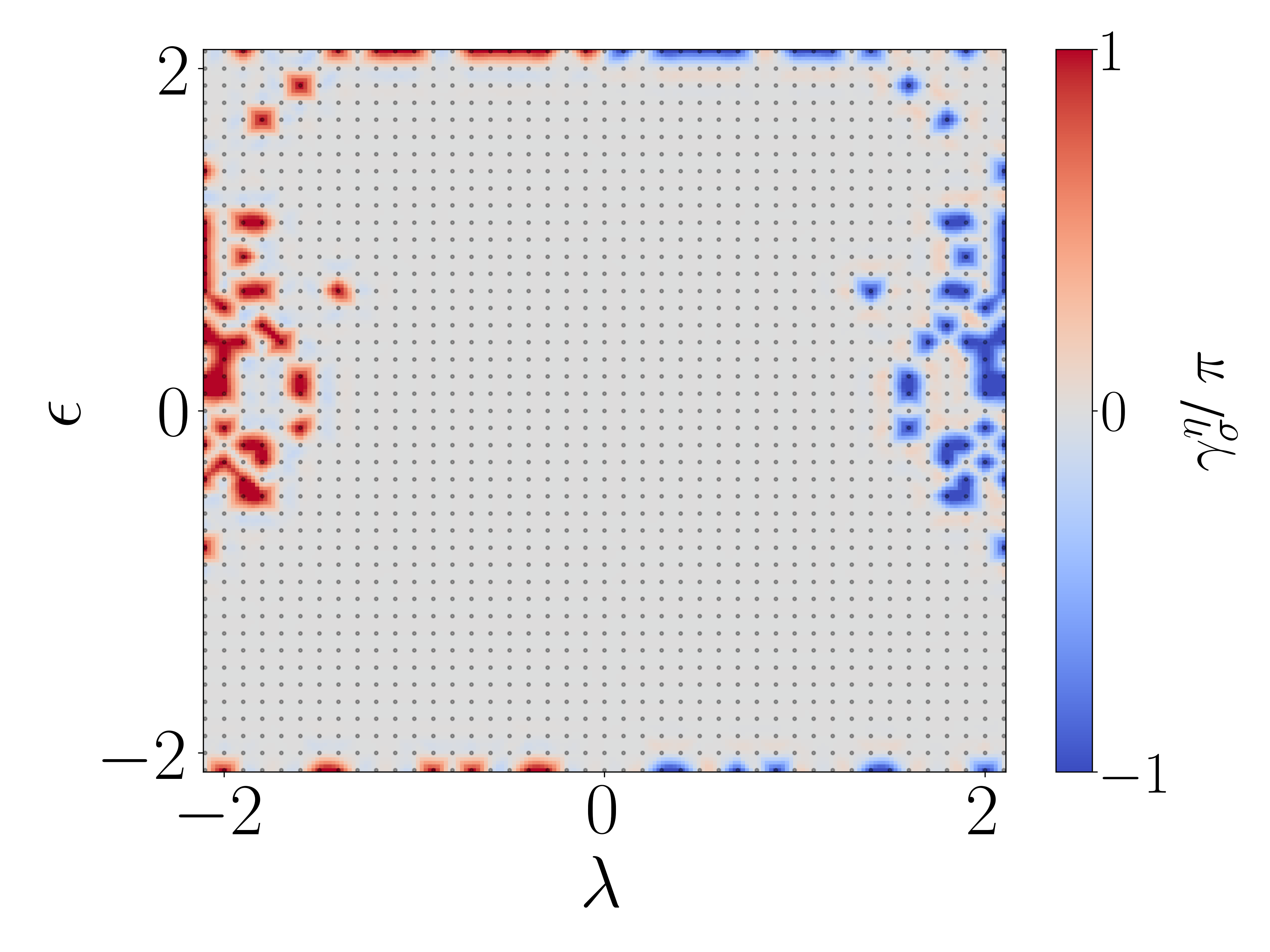}\\
 d) \includegraphics[scale=0.18]{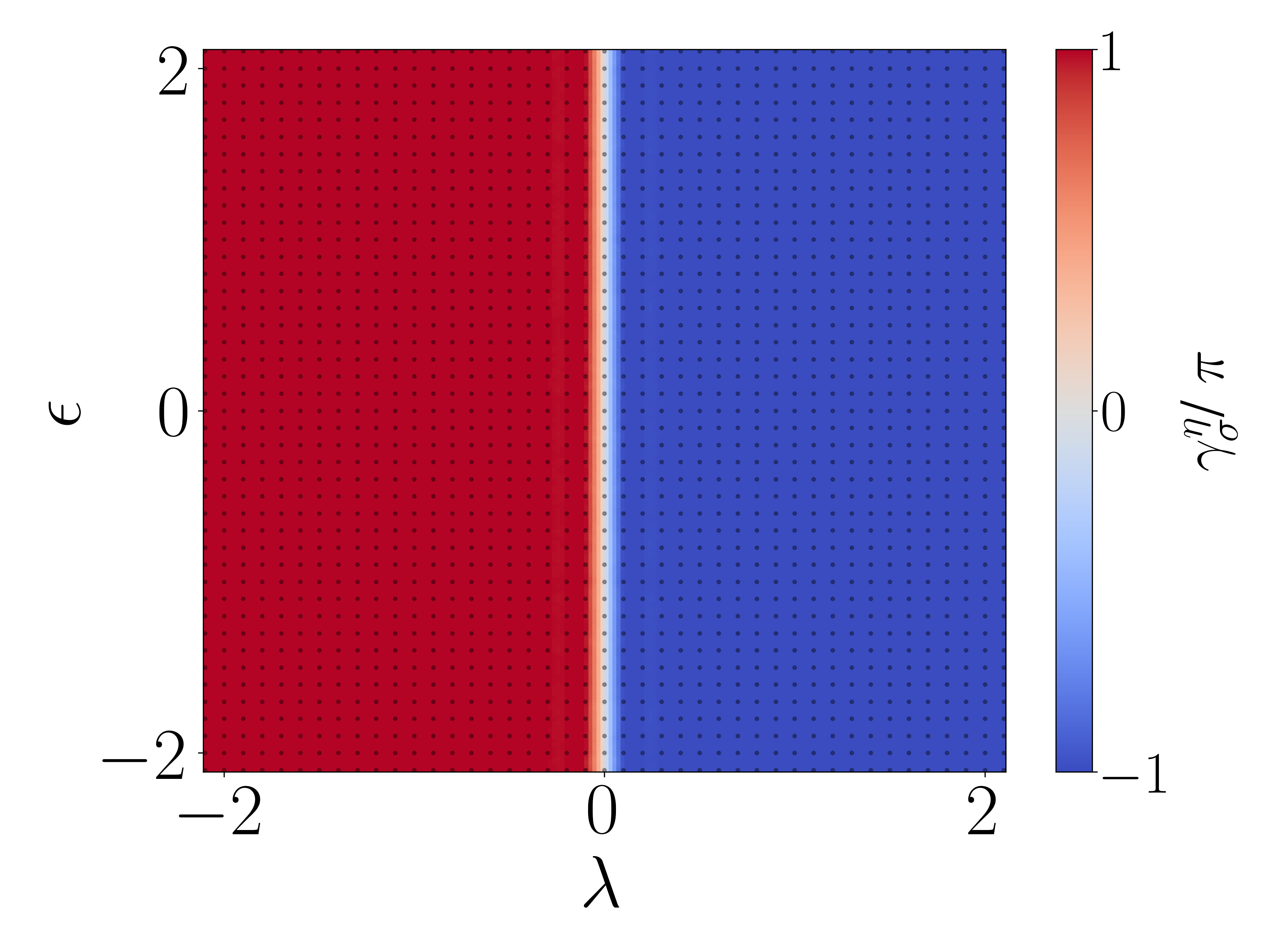}
 e) \includegraphics[scale=0.18]{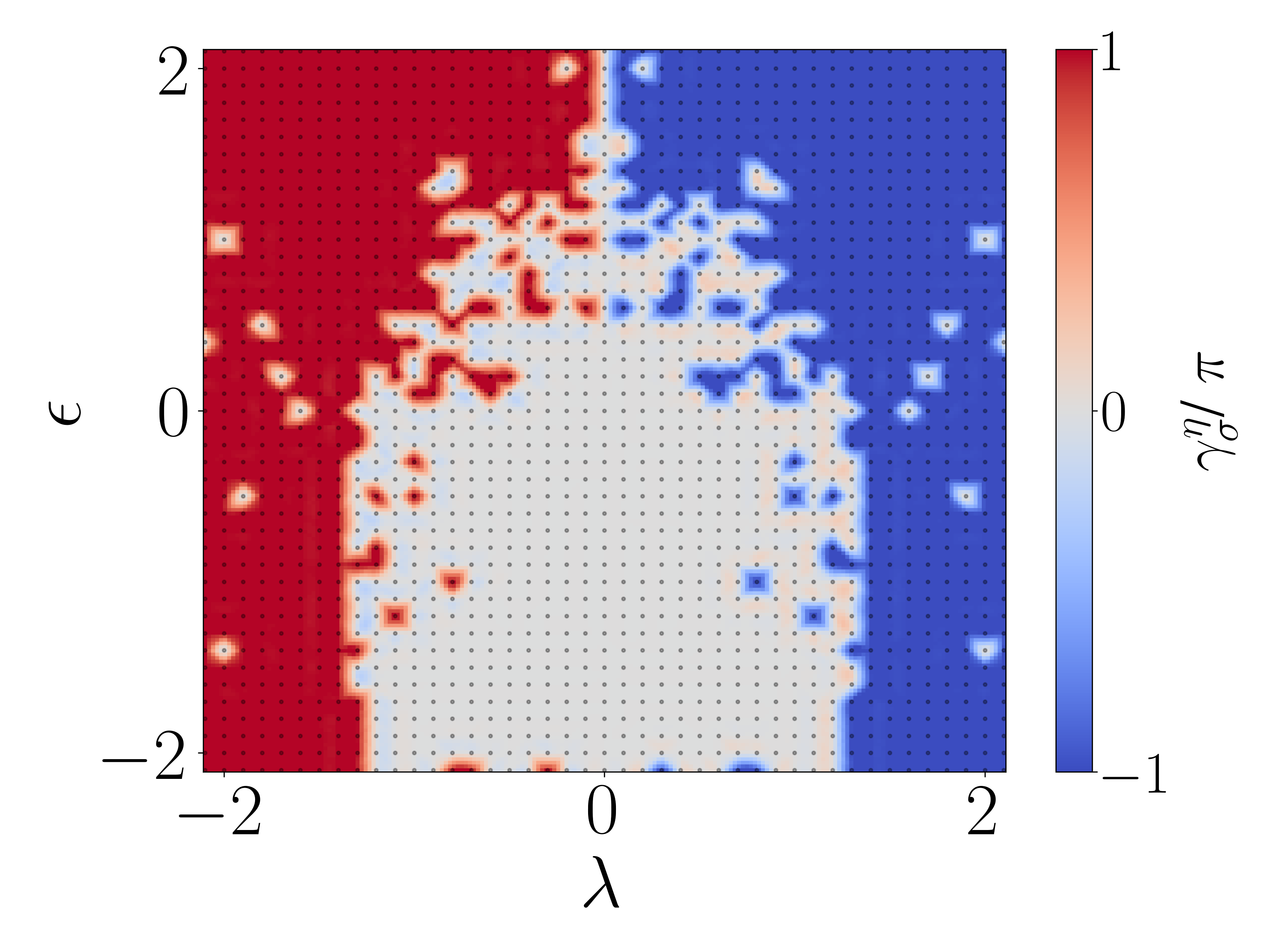}
 f) \includegraphics[scale=0.18]{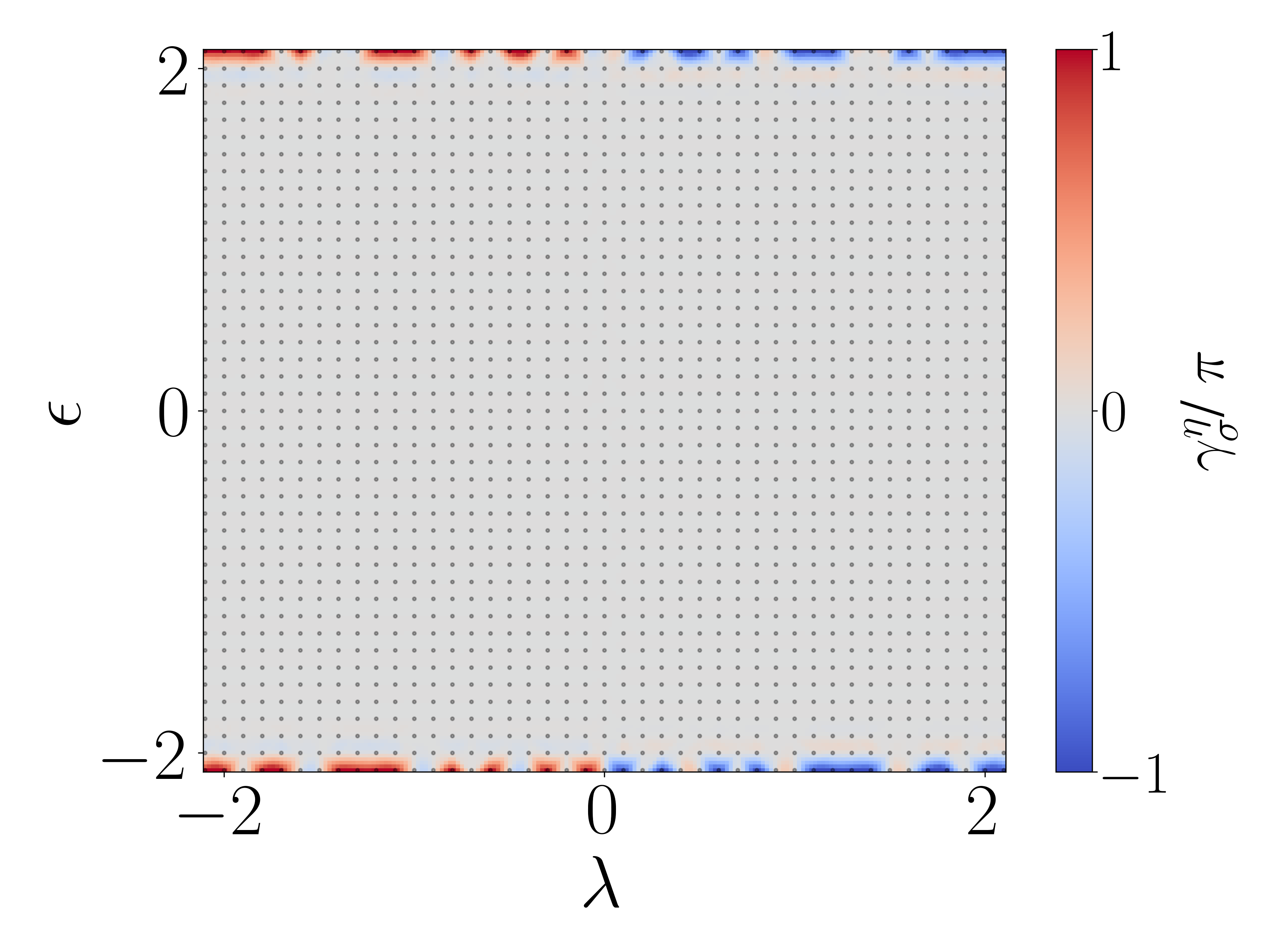}

    \caption{Contour maps of the ground-state Zak phase ($\gamma_{\sigma}^{\eta}/\pi$) (Eq. \ref{eq:Zak_phase}) in the $(\lambda, \bar{\epsilon})$ plane at a fixed temperature  (a-c) $kT=0.25$ eV and (d-f) $kT=1.0$ eV . Panel (a) and (d) shows the result for $U/t_0 = 0.5$, panel (b) and (e)  shows the result for $U/t_0 = 7.5$, while panel (c) and (f) shows the result for $U/t_0 = 10.5$. The topological phase is broken at higher temperatures.}
    \label{fig:contour_maps}
\end{figure*}

Finally, to provide a global view of the topological phases, we constructed contour maps of the ground-state Zak phase in the $(\lambda, \bar{\epsilon})$ plane for fixed values of the interaction $\bar{u}$. As shown in Fig. \ref{fig:contour_maps}, increasing the interaction strength expands the topological region. Comparing different temperatures reveals that thermal effects tend to shrink the size of the topological phase for a given interaction strength, highlighting the competition between interaction-driven ordering and thermal fluctuations.
 

\section{conclusions remarks} \label{sec:conclusions}
In this work, we have presented a comprehensive theoretical and numerical study of the Hubbard model on the Creutz ladder, revealing a rich interplay between electronic correlations, magnetism, and band topology. Using a self-consistent mean-field approach, we have demonstrated that the on-site Hubbard interaction $U$ acts as a powerful tuning parameter capable of inducing a first-order quantum phase transition.

Our primary finding is the identification of a simultaneous magnetic and topological transition. We have shown that as the interaction strength increases, the system's ground state undergoes an abrupt change from an anti-ferromagnetic, topologically non-trivial phase, characterized by a Zak phase of $\pm\pi$, to a ferromagnetic, topologically trivial phase with a Zak phase of $0$. This result establishes a direct and profound connection between the emergence of ferromagnetic order and a distinct topological state, where the spin degrees of freedom and the electronic band structure are intrinsically linked.

By systematically mapping the parameter space, we have constructed detailed phase diagrams that delineate the boundaries between these phases as a function of the on-site energy staggering ($\epsilon$) and the hopping asymmetry ($\lambda$). These diagrams provide a quantitative map of the critical interaction $U_c$ required to drive the transition, showing a complex dependence on the underlying lattice parameters. Our analysis of the energy spectrum was crucial in identifying the first-order nature of this transition, allowing us to distinguish the true ground state from the metastable solutions that coexist beyond the critical point. Furthermore, our investigation of the Fubini-Study metric offers a quantum geometric perspective on the transition. We have confirmed that the topological phase boundary is marked by divergences in the metric, providing a geometric fingerprint of the band-gap closing and reopening that defines the topological change.

These findings establish the Creutz-Hubbard model as a minimal yet powerful platform for exploring interaction-driven topological phenomena. The ability to switch both the magnetic and topological properties of the system with a single parameter ($U$) suggests potential applications in spintronics and quantum information, where such control is highly desirable.

\acknowledgments
 AJEC and GGN thanks the CONAHCyT fellowship (No. CVU 1007044) and the Universidad Nacional Autónoma de México (UNAM) for providing financial support (UNAM DGAPA PAPIIT IN101924 and CONAHCyT project 1564464). We also thanks to Carlos Ernesto L\'opez Natar\'en from Secretaria T\'ecnica de C\'omputo y Telecomunicaciones for his valuable support to implement high-performance numerical calculations. 

\newpage
\bibliographystyle{unsrt}

\begin{thebibliography}{10}
	
	\bibitem{Hasan2010_Review}
	M.~Z. Hasan and C.~L. Kane.
	\newblock Colloquium: Topological insulators.
	\newblock {\em Rev. Mod. Phys.}, 82:3045--3067, Nov 2010.
	
	\bibitem{Tokura2019_Review}
	Yoshinori Tokura, Masashi Kawasaki, and Naoto Nagaosa.
	\newblock Emergent functions of quantum materials.
	\newblock {\em Nature Physics}, 13(11):1056--1068, Nov 2017.
	
	\bibitem{Wen2017_Review}
	Xiao-Gang Wen.
	\newblock Colloquium: Zoo of quantum-topological phases of matter.
	\newblock {\em Rev. Mod. Phys.}, 89:041004, Dec 2017.
	
	\bibitem{Bernevig2013_Book}
	B.~Andrei Bernevig.
	\newblock {\em Topological Insulators and Topological Superconductors}.
	\newblock Princeton University Press, Princeton, 2013.
	
	\bibitem{Kane2005_QSH}
	C.~L. Kane and E.~J. Mele.
	\newblock Quantum spin hall effect in graphene.
	\newblock {\em Phys. Rev. Lett.}, 95:226801, Nov 2005.
	
	\bibitem{Bernevig2006_QSHG}
	B.~Andrei Bernevig, Taylor~L. Hughes, and Shou-Cheng Zhang.
	\newblock Quantum spin hall effect and topological phase transition in hgte
	quantum wells.
	\newblock {\em Science}, 314(5806):1757--1761, 2006.
	
	\bibitem{Espinosa-Champo_2024_Flat_bands}
	Abdiel de~Jesús Espinosa-Champo and Gerardo~G Naumis.
	\newblock Flat bands without twists: periodic holey graphene.
	\newblock {\em Journal of Physics: Condensed Matter}, 36(27):275703, apr 2024.
	
	\bibitem{espinosachampo2025hyperbolicplasmondispersionoptical}
	Abdiel de~Jesús Espinosa-Champo and Gerardo~G. Naumis.
	\newblock Hyperbolic plasmon dispersion and optical conductivity of holey
	graphene: signatures of flat-bands, 2025.
	
	\bibitem{Wang2012_Dirac}
	Zhijun Wang, Yan Sun, Xing-Qiu Chen, Cesare Franchini, Gang Xu, Hongming Weng,
	Xi~Dai, and Zhong Fang.
	\newblock Dirac semimetal and topological phase transitions in ${A}_{3}$bi
	($a=\text{Na}$, k, rb).
	\newblock {\em Phys. Rev. B}, 85:195320, May 2012.
	
	\bibitem{Young2012_Dirac}
	S.~M. Young, S.~Zaheer, J.~C.~Y. Teo, C.~L. Kane, E.~J. Mele, and A.~M. Rappe.
	\newblock Dirac semimetal in three dimensions.
	\newblock {\em Phys. Rev. Lett.}, 108:140405, Apr 2012.
	
	\bibitem{borophene_abdiel}
	Abdiel~E. Champo and Gerardo~G. Naumis.
	\newblock Metal-insulator transition in $8\ensuremath{-}pmmn$ borophene under
	normal incidence of electromagnetic radiation.
	\newblock {\em Phys. Rev. B}, 99:035415, Jan 2019.
	
	\bibitem{Chiu2016_Review}
	Ching-Kai Chiu, Jeffrey C.~Y. Teo, Andreas~P. Schnyder, and Shinsei Ryu.
	\newblock Classification of topological quantum matter with symmetries.
	\newblock {\em Rev. Mod. Phys.}, 88:035005, Aug 2016.
	
	\bibitem{Armitage2018_Weyl}
	N.~P. Armitage, E.~J. Mele, and Ashvin Vishwanath.
	\newblock Weyl and dirac semimetals in three-dimensional solids.
	\newblock {\em Rev. Mod. Phys.}, 90:015001, Jan 2018.
	
	\bibitem{multifractal_abdiel}
	Abdiel Espinosa-Champo and Gerardo~G. Naumis.
	\newblock Multifractal wavefunctions of charge carriers in graphene with folded
	deformations, ripples, or uniaxial flexural modes: Analogies to the quantum
	hall effect under random pseudomagnetic fields.
	\newblock {\em Journal of Vacuum Science \& Technology B}, 39(6):062202, 10
	2021.
	
	\bibitem{Arovas2022_HubbardReview}
	Daniel~P. Arovas, Erez Berg, Steven~A. Kivelson, and Srinivas Raghu.
	\newblock The hubbard model.
	\newblock {\em Annual Review of Condensed Matter Physics}, 13(Volume 13,
	2022):239--274, 2022.
	
	\bibitem{Imada1998_Mott}
	Masatoshi Imada, Atsushi Fujimori, and Yoshinori Tokura.
	\newblock Metal-insulator transitions.
	\newblock {\em Rev. Mod. Phys.}, 70:1039--1263, Oct 1998.
	
	\bibitem{Pesin2010_Mott}
	Dmytro Pesin and Leon Balents.
	\newblock Mott physics and band topology in materials with strong spin--orbit
	interaction.
	\newblock {\em Nature Physics}, 6(5):376--381, May 2010.
	
	\bibitem{Rachel2010_Mott}
	Stephan Rachel and Karyn Le~Hur.
	\newblock Topological insulators and mott physics from the hubbard interaction.
	\newblock {\em Phys. Rev. B}, 82:075106, Aug 2010.
	
	\bibitem{Neupert2011_FCI}
	Titus Neupert, Luiz Santos, Claudio Chamon, and Christopher Mudry.
	\newblock Fractional quantum hall states at zero magnetic field.
	\newblock {\em Phys. Rev. Lett.}, 106:236804, Jun 2011.
	
	\bibitem{Regnault2011_FCI}
	N.~Regnault and B.~Andrei Bernevig.
	\newblock Fractional chern insulator.
	\newblock {\em Phys. Rev. X}, 1:021014, Dec 2011.
	
	\bibitem{Varney2010_InteractionTPT}
	Christopher~N. Varney, Kai Sun, Marcos Rigol, and Victor Galitski.
	\newblock Interaction effects and quantum phase transitions in topological
	insulators.
	\newblock {\em Phys. Rev. B}, 82:115125, Sep 2010.
	
	\bibitem{Budich2012_InteractionTPT}
	Jan~Carl Budich, Ronny Thomale, Gang Li, Manuel Laubach, and Shou-Cheng Zhang.
	\newblock Fluctuation-induced topological quantum phase transitions in quantum
	spin-hall and anomalous-hall insulators.
	\newblock {\em Phys. Rev. B}, 86:201407, Nov 2012.
	
	\bibitem{espinosachampo2025adiabaticitystudytopologicalphases}
	Abdiel de~Jesús Espinosa-Champo, Alejandro Kunold, and Gerardo~G. Naumis.
	\newblock When adiabaticity is not enough to study topological phases in
	solid-state physics: Comparing the berry and aharonov-anandan phases in 2d
	materials, 2025.
	
	\bibitem{Creutz1999}
	Michael Creutz.
	\newblock End states, ladder compounds, and domain-wall fermions.
	\newblock {\em Phys. Rev. Lett.}, 83:2636--2639, Sep 1999.
	
	\bibitem{Leykam2018_FlatBandReview}
	Daniel Leykam, Alexei Andreanov, and Sergej Flach.
	\newblock Artificial flat band systems: from lattice models to experiments.
	\newblock {\em Advances in Physics: X}, 3(1):1473052, 2018.
	
	\bibitem{EspinosaChampo2023_Fubini}
	Abdiel de~Jesús Espinosa-Champo and Gerardo~G Naumis.
	\newblock Fubini–study metric and topological properties of flat band
	electronic states: the case of an atomic chain with s-p orbitals.
	\newblock {\em Journal of Physics: Condensed Matter}, 36(1):015502, sep 2023.
	
	\bibitem{Ryu2010_Classification}
	Shinsei Ryu, Andreas~P Schnyder, Akira Furusaki, and Andreas W~W Ludwig.
	\newblock Topological insulators and superconductors: tenfold way and
	dimensional hierarchy.
	\newblock {\em New Journal of Physics}, 12(6):065010, jun 2010.
	
	\bibitem{Zak1989}
	J.~Zak.
	\newblock Berry's phase for energy bands in solids.
	\newblock {\em Phys. Rev. Lett.}, 62:2747--2750, Jun 1989.
	
	\bibitem{Kang_2020}
	Jin~Hyoun Kang, Jeong~Ho Han, and Y~Shin.
	\newblock Creutz ladder in a resonantly shaken 1d optical lattice.
	\newblock {\em New Journal of Physics}, 22(1):013023, jan 2020.
	
	\bibitem{He2021}
	Yanyan He, Ruosong Mao, Han Cai, Jun-Xiang Zhang, Yongqiang Li, Luqi Yuan,
	Shi-Yao Zhu, and Da-Wei Wang.
	\newblock Flat-band localization in creutz superradiance lattices.
	\newblock {\em Phys. Rev. Lett.}, 126:103601, Mar 2021.
	
	\bibitem{Meng2016}
	Zengming Meng, Lianghui Huang, Peng Peng, Donghao Li, Liangchao Chen, Yong Xu,
	Chuanwei Zhang, Pengjun Wang, and Jing Zhang.
	\newblock Experimental observation of a topological band gap opening in
	ultracold fermi gases with two-dimensional spin-orbit coupling.
	\newblock {\em Phys. Rev. Lett.}, 117:235304, Dec 2016.
	
	\bibitem{Rechtsman2013_Photonic}
	Mikael~C. Rechtsman, Julia~M. Zeuner, Yonatan Plotnik, Yaakov Lumer, Daniel
	Podolsky, Felix Dreisow, Stefan Nolte, Mordechai Segev, and Alexander
	Szameit.
	\newblock Photonic floquet topological insulators.
	\newblock {\em Nature}, 496(7444):196--200, Apr 2013.
	
	\bibitem{Mukherjee2017_PhotonicCreutz}
	Sebabrata Mukherjee, Alexander Spracklen, Manuel Valiente, Erika Andersson,
	Patrik {\"O}hberg, Nathan Goldman, and Robert~R. Thomson.
	\newblock Experimental observation of anomalous topological edge modes in a
	slowly driven photonic lattice.
	\newblock {\em Nature Communications}, 8(1):13918, Jan 2017.
	
	\bibitem{Piga2017}
	J.~J\"unemann, A.~Piga, S.-J. Ran, M.~Lewenstein, M.~Rizzi, and A.~Bermudez.
	\newblock Exploring interacting topological insulators with ultracold atoms:
	The synthetic creutz-hubbard model.
	\newblock {\em Phys. Rev. X}, 7:031057, Sep 2017.
	
	\bibitem{Kuno_2020}
	Yoshihito Kuno, Takahiro Orito, and Ikuo Ichinose.
	\newblock Flat-band many-body localization and ergodicity breaking in the
	creutz ladder.
	\newblock {\em New Journal of Physics}, 22(1):013032, jan 2020.
	
	\bibitem{Atala2013_ColdAtom}
	Marcos Atala, Monika Aidelsburger, Julio~T. Barreiro, Dmitry Abanin, Takuya
	Kitagawa, Eugene Demler, and Immanuel Bloch.
	\newblock Direct measurement of the zak phase in topological bloch bands.
	\newblock {\em Nature Physics}, 9(12):795--800, Dec 2013.
	
	\bibitem{Aidelsburger2013_ColdAtom}
	M.~Aidelsburger, M.~Atala, M.~Lohse, J.~T. Barreiro, B.~Paredes, and I.~Bloch.
	\newblock Realization of the hofstadter hamiltonian with ultracold atoms in
	optical lattices.
	\newblock {\em Phys. Rev. Lett.}, 111:185301, Oct 2013.
	
	\bibitem{Zurita2020}
	Juan Zurita, Charles~E. Creffield, and Gloria Platero.
	\newblock Topology and interactions in the photonic creutz and creutz-hubbard
	ladders.
	\newblock {\em Advanced Quantum Technologies}, 3(2):1900105, 2020.
	
	\bibitem{Bouzerar2025}
	Maxime Thumin and Georges Bouzerar.
	\newblock {Correlation functions and characteristic lengthscales in flat band
		superconductors}.
	\newblock {\em SciPost Phys.}, 18:025, 2025.
	
	\bibitem{Neel1948_Antiferromagnetism}
	L.~N{\'e}el.
	\newblock Propri{\'e}t{\'e}s magn{\'e}tiques des ferrites; ferrimagn{\'e}tisme
	et antiferromagn{\'e}tisme.
	\newblock {\em Annales de Physique}, 12(3):137--198, 1948.
	
	\bibitem{Stoner1938_Ferromagnetism}
	Edmund~Clifton Stoner.
	\newblock Collective electron ferromagnetism.
	\newblock {\em Proceedings of the Royal Society of London. Series A.
		Mathematical and Physical Sciences}, 165(922):372--414, 1938.
	
	\bibitem{Provost1980_FSMetric}
	J.~P. Provost and G.~Vallee.
	\newblock Riemannian structure on manifolds of quantum states.
	\newblock {\em Communications in Mathematical Physics}, 76(3):289--301, Sep
	1980.
	
	\bibitem{Espinosa-Champo_2024}
	Abdiel de~Jesús Espinosa-Champo and Gerardo~G Naumis.
	\newblock Fubini–study metric and topological properties of flat band
	electronic states: the case of an atomic chain with s-p orbitals.
	\newblock {\em Journal of Physics: Condensed Matter}, 36(1):015502, sep 2023.
	
	\bibitem{Leumer_2020}
	Nico Leumer, Magdalena Marganska, Bhaskaran Muralidharan, and Milena Grifoni.
	\newblock Exact eigenvectors and eigenvalues of the finite kitaev chain and its
	topological properties.
	\newblock {\em Journal of Physics: Condensed Matter}, 32(44):445502, aug 2020.
	
	\bibitem{verma2025}
	Nishchhal Verma, Philip J.~W. Moll, Tobias Holder, and Raquel Queiroz.
	\newblock Quantum geometry: Revisiting electronic scales in quantum matter,
	2025.
	
	\bibitem{Komissarov2024}
	Ilia Komissarov, Tobias Holder, and Raquel Queiroz.
	\newblock The quantum geometric origin of capacitance in insulators.
	\newblock {\em Nature Communications}, 15(1):4621, May 2024.
	
	\bibitem{yu2025}
	Jiabin Yu, B.~Andrei Bernevig, Raquel Queiroz, Enrico Rossi, Päivi Törmä,
	and Bohm-Jung Yang.
	\newblock Quantum geometry in quantum materials, 2025.
	
	\bibitem{TomokiOzawa2018}
	Tomoki Ozawa and Nathan Goldman.
	\newblock Extracting the quantum metric tensor through periodic driving.
	\newblock {\em Phys. Rev. B}, 97:201117, May 2018.
	
	\bibitem{Aleksi2021}
	Aleksi Julku, Georg~M. Bruun, and P\"aivi T\"orm\"a.
	\newblock Excitations of a bose-einstein condensate and the quantum geometry of
	a flat band.
	\newblock {\em Phys. Rev. B}, 104:144507, Oct 2021.
	
	\bibitem{Cayssol_2021}
	J~Cayssol and J~N Fuchs.
	\newblock Topological and geometrical aspects of band theory.
	\newblock {\em Journal of Physics: Materials}, 4(3):034007, apr 2021.
	
	\bibitem{Kruchkov2022}
	Alexander Kruchkov.
	\newblock Quantum geometry, flat chern bands, and wannier orbital quantization.
	\newblock {\em Phys. Rev. B}, 105:L241102, Jun 2022.
	
	\bibitem{BernevigBogdan2022}
	Kukka-Emilia Huhtinen, Jonah Herzog-Arbeitman, Aaron Chew, Bogdan~A. Bernevig,
	and P\"aivi T\"orm\"a.
	\newblock Revisiting flat band superconductivity: Dependence on minimal quantum
	metric and band touchings.
	\newblock {\em Phys. Rev. B}, 106:014518, Jul 2022.
	
	\bibitem{TomokiOzawa2019}
	Tomoki Ozawa and Nathan Goldman.
	\newblock Probing localization and quantum geometry by spectroscopy.
	\newblock {\em Phys. Rev. Research}, 1:032019, Nov 2019.
	
	\bibitem{bengtsson_zyczkowski_2006}
	Ingemar Bengtsson and Karol Zyczkowski.
	\newblock {\em Geometry of Quantum States: An Introduction to Quantum
		Entanglement}.
	\newblock Cambridge University Press, 2006.
	
	\bibitem{zerounian2020sisl}
	Nick~R. Zerounian, Nick Papior, and Mads Brandbyge.
	\newblock sisl: v0.9.6, 2020.
	
	\bibitem{sanzwuhl2023hubbard}
	Sofia Sanz~Wuhl, Nick Papior, Mads Brandbyge, and Thomas Frederiksen.
	\newblock hubbard: A python package for hubbard model calculations, 2023.
	\newblock v0.1.0.
	
	\bibitem{Asboth2016_Book}
	J{\'a}nos~K. Asb{\'o}th, L{\'a}szl{\'o} Oroszl{\'a}ny, and Andr{\'a}s
	P{\'a}lyi.
	\newblock {\em A Short Course on Topological Insulators}, volume 919 of {\em
		Lecture Notes in Physics}.
	\newblock Springer International Publishing, 2016.
	
\end{thebibliography}

\end{document}